\documentclass[useAMS,usenatbib]{mn2e}
\usepackage{epsfig,amsmath,amssymb,amsfonts,mathrsfs,latexsym,graphicx}
\begin{document}

\title[Oxygen Recombination in CC-SNe]{Oxygen Recombination in the
nebular phase of supernovae 1998bw and 2002ap$^{*}$}

\author[Maurer et al.]{I. Maurer$^{1,**}$, P. A. Mazzali$^{1,2,3}$
\\$^1$ Max Planck Institut f\"ur
  Astrophysik, Karl-Schwarzschild-Str.1, 85741 Garching, Germany
\\$^2$ Scuola Normale Superiore, Piazza dei Cavalieri, 7, 56126 Pisa, Italy
\\$^3$ National Institute for Astrophysics-OAPd, Vicolo dell'Osservatorio, 5, 35122 Padova, Italy
\\
\\$^{**} maurer@mpa$-$garching.mpg.de$
}

\maketitle

\begin{abstract}
Late-time spectra of stripped-envelope CC-SNe are dominated by strong
[O {\sc i}] $\lambda\lambda$6300,6363 emission, caused by thermal
electron excitation of forbidden [O {\sc i}] transitions. The permitted O
  {\sc i} 7774~\AA\ line is also often observed. This line cannot result from thermal
electron excitation of the oxygen ground state. In this work tests are performed to verify whether the line can be
powered by oxygen recombination alone, using the examples of two of the best
studied type Ic SNe, 1998bw and 2002ap. 

Temperature-dependent effective recombination coefficients for neutral
oxygen are calculated using
available atomic data. Missing atomic data are computed in a temperature range typical for SN
nebulae. Core ejecta models for SNe 1998bw and 2002ap are obtained from modelling their
nebular emission spectra so that oxygen recombination line formation
is computed consistently with
oxygen forbidden line emission.

While SN 2002ap can be explained well by a one
dimensional shell model, this seems not to be possible for SN
1998bw, for which a two dimensional model is
found. At very late epochs the formation of the
O {\sc i} 7774~\AA\ line can be explained by recombination
radiation for both SNe, but at earlier epochs strong absorption is
present which may determine the
strength of this line even at $\sim$ 200 days.

\end{abstract}


\maketitle

\section{Introduction}
\label{intro}

Massive stars ($>8$ M$_\odot$) collapse when the nuclear fuel in their
central regions is consumed, producing a core-collapse supernova (CC-SN) and
forming a black hole or a neutron star. CC-SNe with a H-rich spectrum
are classified as Type II \citep{Filippenko97}. If the star was
stripped of at least most of its H envelope prior to the explosion, the
SNe are classified according to the degree of stripping as Type IIb (strong He lines, and weak but clear H), Type Ib (strong
He lines but no H), and Type Ic (no He or H lines).

Some CC-SNe, called broad-lined SNe (BL-SNe), exhibit very broad
absorption lines at early times, caused by the presence of sufficiently massive ejecta
expanding at high velocities. These SNe are sometimes called hypernovae, and can be
associated with long-duration gamma-ray bursts (GRBs) \citep[see][and references
therein]{Woosley06}. 

In GRB scenarios a relativistic outflow launched by a central engine
deposits some fraction of its energy into the SN ejecta. Since the
energy is probably deposited preferentially
along the polar axis, the SN might be strongly asymmetric
\citep[e.g.,][]{Maeda02}. The nearest, best-studied GRB-SNe are SN
1998bw / GRB 980425 \citep{Galama99}, SN 2003dh / GRB 030329 
\citep{Matheson04}, SN
2003lw / GRB 031203 \citep{Malesani04}, and SN 2006aj / GRB/XRF 060218
\citep{Pian06}. It is not yet fully established whether the GRBs
(or X-ray flashes) accompanying nearby CC-SNe share the same
properties of 
high-redshift GRBs. CC-SNe may also be characterised by asphericities, although a jet 
does not necessarily form
\citep{Blondin03,Kotake04,Moiseenko06,Burrows07,Takiwaki09}.

Asphericities in the inner and outer ejecta 
are evident in at least some CC-SNe. Two clear indicators are
velocity differences of Fe and lighter-element lines \citep[e.g.,][]{Mazzali01}, and polarisation
measurements \citep[e.g.,][]{Hoflich91}. Indirect
indications also emerge from a comparison of the inner and outer ejecta velocities \citep{Maurer09}.

Independent of their type, SNe become increasingly transparent to
optical light with time, as the ejecta thin out. At late times ($> 200$ days after the
explosion), the innermost layers of the SN can be observed. This epoch is called
the nebular phase, because the spectrum turns from being dominated by
absorption to an
emission spectrum, mostly showing forbidden lines.  In this phase the radiated energy of a SN
is provided by the decay of radioactive $^{56}$Co (which is produced by the earlier
decay of $^{56}$Ni) into $^{56}$Fe. Decaying $^{56}$Co emits $\gamma$-rays and positrons which
are absorbed by the SN ejecta. As the deposition rate of $\gamma$-rays and
positrons depends on the density and $^{56}$Ni distribution, the inner parts of the
SN dominate the nebular spectra. Therefore the nebular phase is especially suitable for studying the core of SNe.

Several authors have modelled nebular-phase spectra of SNe to derive quantities 
such as the $^{56}$Ni mass
\citep[e.g.,][]{Mazzali04,Stritzinger06,Sauer06,Maeda07},
ejecta velocities \citep[e.g.,][]{Mazzali07,Taubenberger09,Maurer09}, asphericities
\citep[e.g.,][]{Mazzali05,Maeda06,Mazzali07b,Mazzali08}, and
elemental abundances \citep[e.g.,][]{Maeda07, Maeda07b}.

To infer SN core ejecta velocities and geometry usually the [O {\sc i}]
$\lambda\lambda$ 6300, 6364 doublet is investigated, since this is by
far the strongest nebular emission line (SNe of type Ic), especially at very late epochs. These lines are formed by
thermal electron excitation of the 2p($^1$D) state. Higher quantum states
(n $\ge$ 3) cannot be excited by thermal electrons sufficiently
because of
the large ratio of excitation energy ($\sim$ 9~eV) and electron
temperature ($\sim 0.4$~eV). Possible excitation mechanisms are
absorption, non-thermal electrons and recombination.

Previous studies of central oxygen have focused almost exclusively on the [O {\sc i}]
$\lambda\lambda$6300, 6364 doublet.  It is useful to investigate whether
extending the analysis to other oxygen lines gives a consistent
picture. Therefore it is important to
understand the formation of these lines in detail.

In Section 2 
temperature-dependent effective recombination rates are obtained for the n = 3 levels of
neutral oxygen and some quantities relevant for oxygen
recombination and emission-line formation in CC-SN nebulae are
estimated. A one dimensional shell model for SN 2002ap is described in
Section 3 and a two-dimensional model for SN 1998bw in Section 4 emphasising the role of the O {\sc i} 7774~\AA\ line. Results are discussed in Section 5.

\section{Oxygen lines in the nebular phase of CC-SNe}
\label{oxrec}

\begin{figure} 
\begin{center}
\includegraphics[width=8.5cm, clip]{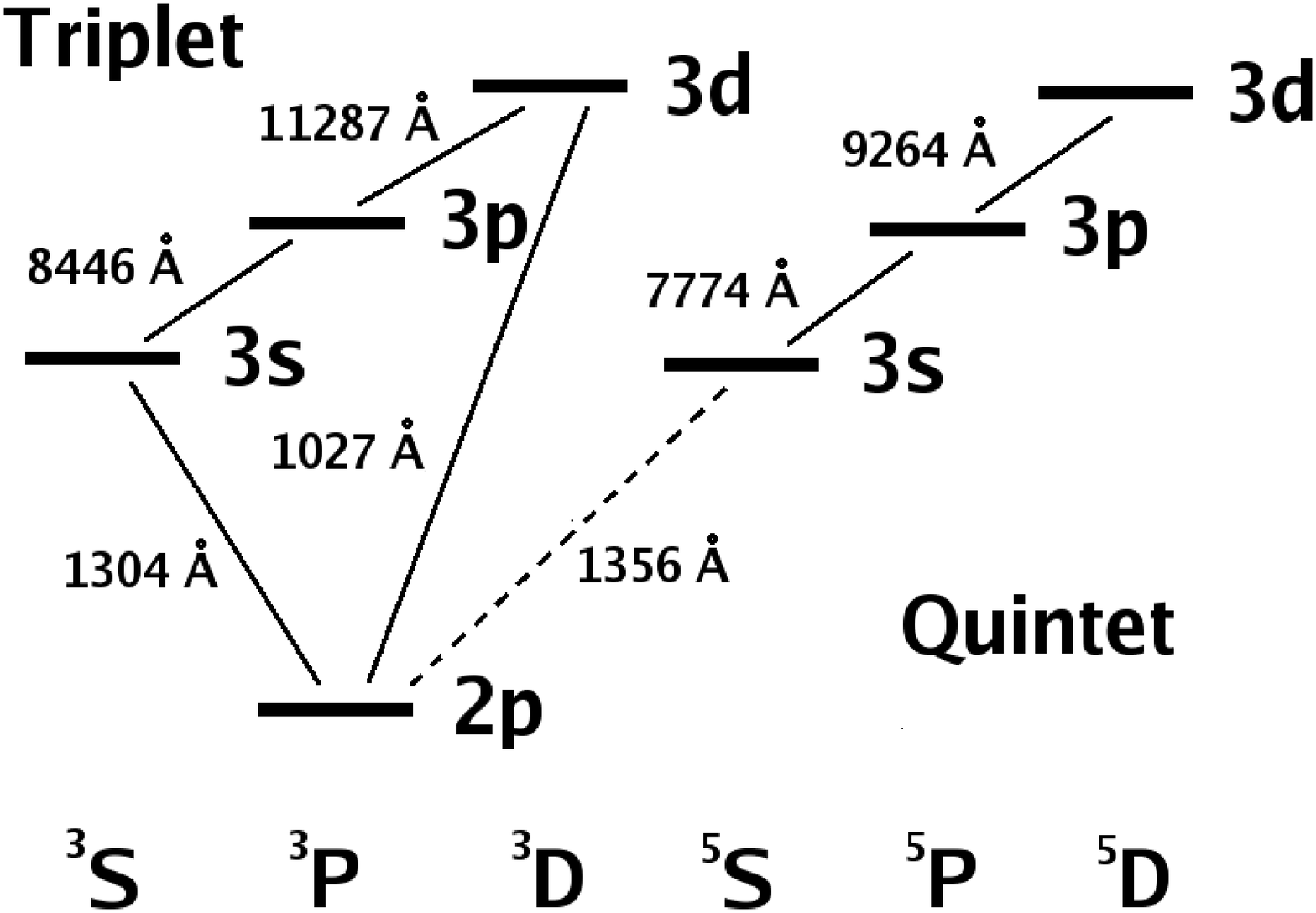}
\end{center}
\caption{Neutral oxygen levels that are
  potentially interesting for the nebular phase of CC-SNe. The
  singlet states (which are responsible for the nebular oxygen forbidden-line emission)
  and states with n $\ge$ 4  are not shown. The 2p, 3p and 3d states can be
  subdivided into several j sub-states, which are not shown to keep the
  illustration clear. The wavelengths given represent the
  mean wavelengths of emission. Lines emitted by the different j sub-states differ
  slightly from the values given. For example, the line we refer to as O {\sc
    i} 7774~\AA\ in this paper consists of three lines at wavelengths
  7771.94, 7774.17 and 7775.39~\AA . While all transitions indicated by
  solid lines are permitted with radiative rates of order 10$^{7-8}$ s$^{-1}$, the 3s($^5$S) to
2p($^3$P) transition (spin-forbidden; dashed line) has a radiative rate of order
10$^3$ s$^{-1}$ only.} 
\label{oxgrot}
\end{figure}

In this section temperature-dependent effective
recombination rates are obtained for  the 3s, 3p
and 3d triplet and quintet states of neutral oxygen. These rates
are used in our nebular code to calculate oxygen
recombination line emission. These levels are responsible for O {\sc i} 7774,
8446, 9264 and 11287 A line emission (see Figure \ref{oxgrot} for
illustration). We further investigate the relevant
processes for oxygen recombination and absorption line formation in
the nebular phase of CC-SNe. 

Recombination into the 2p singlet states is not important in the CC-SN
nebular phase. About
80 $-$ 90$\%$ of the radioactive energy is deposited in thermal electrons,
while only 10 $-$ 20\% goes into ionisation (see below). Since
the 2p singlet states can be efficiently excited by thermal electron
collisions, the population of these levels is predominantly determined
by this process. 

\subsection{Effective recombination rates for neutral oxygen}
\label{effrec}

\begin{figure} 
\begin{center}
\includegraphics[width=8.5cm, clip]{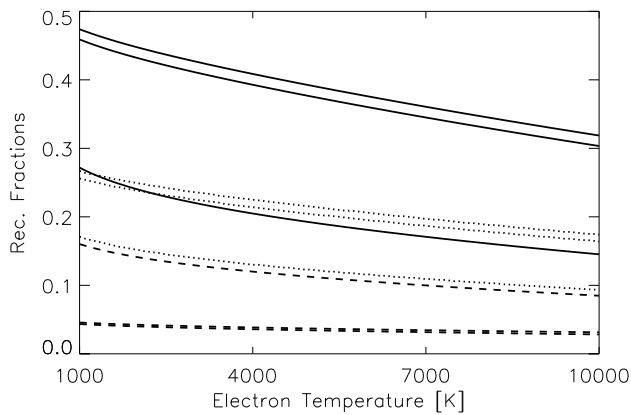}
\end{center}
\caption{Temperature-dependent normalised effective recombination
  fractions. The quintet states (3d, 3p, 3s from bottom to top) are
  shown by solid lines. Triplet states (3p, 3s, 3d from bottom
  to top; note that 3p and 3s do overlap) for the case of optically thin
  ground state transitions are shown by dashed lines.
   Triplet states (3d, 3p, 3s from bottom
  to top) for the case of optically thick ground state transitions are shown by dotted lines.}
\label{effrecfrac}
\end{figure}

Di-electronic recombination is much weaker than radiative
recombination [roughly a factor
of 5 \citep{Nussbaumer83} at 10000~K and decreasing rapidly with temperature]
and is not included when calculating the recombination
fractions. At high electron densities collisional transitions between
excited states can become important. However, for the electron
densities expected in the nebular
phase ($n_\mathrm{e} < 10^{10}$ cm$^{-3}$), this effect is weak. At high electron densities ($n_\mathrm{e}
\sim 10^{10}$ cm$^{-3}$) and low temperatures ($T \le 3000$K) collisional recombination can become important
\citep[e.g. see][]{Storey95}. However, during the nebular phase only the
combinations high density, high temperature (early) and low
density, low temperature (late) are important and the contribution of
collisional recombination is weak (e.g. $\sim 20\%$ at $n_\mathrm{e} \sim
10^{9}$ cm$^{-3}$ and $T \sim 5000$K).

Therefore we only consider radiative recombination.
We are interested in which fraction of
recombining electrons reaches a certain atomic level, whether by direct
recombination or cascading from higher states. It had been noted by
\citet{Julienne74,Sunggi91} that high quantum
levels (n $\sim$ 20) might be important to calculate the recombination cascade.
In the literature recombination coefficients and radiative rates are
available for neutral oxygen up to maximum quantum numbers n $\sim$ 10
(see e.g. TIPTOPbase \footnote{http://cdsweb.u-strasbg.fr/topbase/},
NIST \footnote{http://www.nist.gov/index.html}).

For azimuthal quantum numbers l $\ge$ 3 one can use the hydrogenic approximation,
however for the s, p and d orbitals this should be avoided
\citep[e.g.][]{Sunggi91}.  Therefore, we obtained radiative recombination coefficients and radiative
rates for all s, p and d levels of oxygen up to n = 20 using a quantum
defect method (QDM)
\citep{Bates49,Seaton58,Burgess60}. 

The potential of the atomic core and
the inner electrons approaches the asymptotic form $1/r$ quickly. The
wave functions of the excited states can be described by a hydrogen-like
solution, with the difference that the quantum energy levels are shifted
owing to the unknown structure of the core potential. This shift of
energy levels is described by the quantum defect (defined as the difference between the real and 'effective'
principal quantum number) and can be measured in experiments or is determined by more
sophisticated calculations.

The wave functions calculated by the QDM are described with sufficient accuracy
at radii larger than a few (atomic units) and therefore the QDM is reliable as long as the
main contribution of the transition integrals comes from sufficiently
large radii and there is no strong cancellation in the integrand.
Especially for ground state transitions, where contributions from
small radii are important, the QDM is therefore not reliable.

To improve the calculation, we numerically integrate the electron wave equation \citep[e.g.][]{Seaton58}
\begin{equation}
\Big[\frac{d^2}{dr^2} - \frac{l(l+1)}{r^2} - V(r) + E\Big]P(E,l;r) = 0
\end{equation}
inwards, starting with the solution obtained by the QDM at large radii and using a Thomas-Fermi
potential 
\begin{equation}
V(r) = \frac{2N}{r}\exp(-Z^{1/3}r) + \frac{2(Z - N)}{r}
\end{equation}
where Z is the charge of the nucleus and N the number of core
electrons. This integration is not part of the QDM. The Thomas-Fermi potential gives the correct asymptotic
behaviour of the atomic potential for  $r \rightarrow 0$ and  $r
\rightarrow \infty$ \citep{Burke75}. The result of this integration is a behaviour of the
wave-function for r $\rightarrow$ 0 different than predicted by the
QDM. For most transitions the contribution from small radii is
approximately zero and therefore the calculated rates are not
influenced by this correction. For the transitions which are
influenced by the inner parts of the wave function  we
achieve better agreement with oxygen atomic data provided in the
literature at low quantum numbers (n $<$ 10).

The difference
for ground state radiative transitions relative to NIST recommended data
is always less than 50$\%$ and the agreement increases considerably with the quantum
number of the excited states. For ground state transitions from
levels n $ >$ 5  the disagreement becomes less than 10$\%$. For
transitions between excited states the maximum difference to NIST
recommended data is 40$\%$ with most transitions agreeing at the 10$\%$
level or better. Whenever the disagreement was worse than 10$\%$ we
replaced our data with the values from the literature. We
compared our direct and effective recombination coefficients to \citet{Julienne74,Sunggi91} and found
good agreement (better than 10$\%$), as we did for the
effective recombination rates at 1160~K. Our total recombination rate agrees
with the value given by \citet{Sunggi91} to 1 $\%$ at 10000~K. However compared
to \citet{Aldrovandi73} our total recombination rate is $\sim$ 25$\%$
too low. It was noted by \citet{Julienne74}
that neglecting the n $> 20$ levels might cause an underestimate of
$\sim$ 20$\%$ of the total recombination rate (at temperatures around
1000~K; the effect becomes smaller at higher temperatures). More
importantly \citet{Aldrovandi73} used a hydrogenic approximation for
all atomic states, which leads to an overestimate of the recombination
rate \citep{Sunggi91}. A comparison to a recent
calculation from \citet{Badnell06} shows that our total radiative
recombination rates agree to their results within 15\% . To handle
that deviation, we normalise our result.

We calculate the ratio of the effective recombination rates of the n =
3 levels to the total recombination rate into excited states and find
that this ratio varies only weakly with the number of levels
included (it changes by $\sim$ 1$\%$ (3s), 1$\%$ (3p) and 8$\%$ (3d) when using a maximum quantum number of n = 20 instead of n =
10) and is not too sensitive to uncertainties in the atomic data
(see below), while the total recombination rate is (it increases
by 15 - 30$\%$ depending on temperature when using a maximum quantum
number n = 20 instead of n = 10).

Since fitting formulae for the total and ground state recombination
rates are available in the literature, we can obtain the total
recombination rate into excited states and
normalise our effective recombination rates to the total radiative recombination
rate. We call this the normalised effective recombination fraction.

The recombination rates used for this normalisation are the total
recombination rate of \citet{Badnell06}
\begin{equation}
R_\mathrm{O[I]} = A\Bigg[\sqrt{\frac{T}{T_0}}\Big(1 +
  \frac{T}{T_0}\Big)^{1 - B'}\Big(1 + \frac{T}{T_1}\Big)^{1 +B'}\Bigg]^{-1} \mathrm{cm^3 s^{-1}}
\end{equation} 
with $B' = B + C\exp\big(-\frac{T_2}{T}\big)$, $A  = 6.622 \cdot
10^{-11}$, $B  = 0.6109$, $C  = 0.4093$, $T_0  = 4.136$, $T_1  = 4.214
\cdot 10^{6}$ and $T_2  = 8.770 \cdot 10^{4}$ and the (direct) ground state recombination rate of \citet{Pequignot90}
\begin{equation}
\begin{split}
R_\mathrm{GS} & = (1.174 + 0.2463x + 0.2144x^2 - 0.0621x^3)\\
 & \times 10^{-13} T_4^{-0.5} \mathrm{cm^3 s^{-1}}
\end{split}
\end{equation}
with $x = \log(1 + T_4)/\log(2)$ and $T_4 = T/10000$~K. These normalised
effective recombination fractions are shown in Figure \ref{effrecfrac}
and are listed in Appendix~B. To obtain absolute values one has to
multiply by the total radiative recombination rate. Since the total
recombination rate contains a contribution from direct ground state
recombination and since lower levels contain cascading contributions
from higher states, these fractions do not add up to one.

Since there are uncertainties in our atomic data we want to quantify
their influence. We randomly vary all the radiative rates obtained by the QDM by a value between $\pm$
20$\%$ and the recombination rates obtained by the QDM by a value
between $\pm$ 50$\%$ several times. This should be larger than the typical error, at least for
excited states. The resulting deviations from the standard values are
typically a few percent, often $\ll$ 5$\%$. 

\subsection{Recombination line formation}
\label{reclf}

\begin{figure} 
\begin{center}
\includegraphics[width=8.5cm, clip]{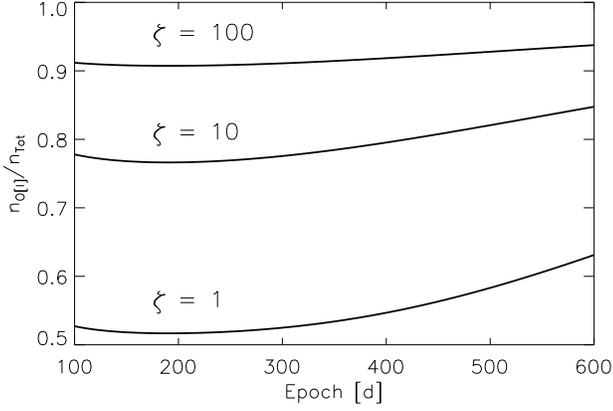}
\end{center}
\caption{The ratio
 $n_\mathrm{O[I]}/n_\mathrm{Tot}$ for different clumping factors
  ($\zeta =$ 1, 10, 100 from bottom to top) calculated for the
    test model described in Appendix A. Since clumping increases
    recombination the fraction of neutral oxygen increases with $\zeta$.} 
\label{pap4test4}
\end{figure}

With the effective recombination rates shown in Figure \ref{effrecfrac} it
is possible to calculate the strength of all [O {\sc i}] n = 3
lines due to recombining electrons. 

The population of excited oxygen levels by recombination is
implemented in the one dimensional \citep{Mazzali01,Mazzali07b} and three dimensional
version \citep{Maurer09} of our nebular code and will be used in
Sections \ref{2002ap} and \ref{1998bw} to calculate synthetic recombination
lines for SNe  1998bw and 2002ap.

However, to obtain insight into the formation of these lines it is useful to derive some estimates for the recombination line
formation. Below, we estimate the strength of the O {\sc i}
7774~\AA\ line. This is the
strongest non-ground state (NGS) recombination line thanks to
its high effective recombination rate and transition energy [3p($^5$P)
  to 3s($^5$S)]. Ground state recombination line observations are usually not
available since most spectra reach a minimum wavelength of 3000~\AA\ $-$ 4000~\AA\ only.

The ionisation rate can be calculated using the concept of 'work
per ion' \citep[e.g.][]{Axelrod80}, which describes the energy lost 
to thermal electrons in order to produce one ionisation. The
ionisation rate per atom and ion is
then given by 
\begin{equation}
Y_\mathrm{O[I]} = \frac{L_\mathrm{Dep}}{N_\mathrm{Tot}W_\mathrm{O[I]}}
\end{equation}
where $L_\mathrm{Dep}$ is the total deposited luminosity, $N_\mathrm{Tot}$ is the
total number of atoms and $W_{O[I]}$ is the energy lost to thermal
electrons per ionisation and can be estimated by 
comparing the ionisation cross-section with atomic and plasma loss
functions (more details are given in Section \ref{ion}).

During the first several hundred days of a SN, ionisation
balance is a valid assumption \citep{Axelrod80}, therefore
\begin{equation}
Y_\mathrm{O[I]}(1 + {\cal R})n_\mathrm{O[I]} =
Y'_\mathrm{O[I]}n_\mathrm{O[I]} =
R_\mathrm{O[I]}n_\mathrm{e}\zeta n_\mathrm{O[II]}
\end{equation}
where ${\cal R}$ is the recycling fraction \citep[recombination radiation causes
further ionisation and ${\cal R}$ is expected to have a value
between 0.3 and 0.5;][]{Axelrod80} and $\zeta$ is the clumping
factor (explained in more detail below), defined as the inverse of the filling factor \citep{Li92}.

Since oxygen is the dominant element in CC-SNe cores [$\sim$ 55 \% in SN
2002ap and $\sim$ 75 \% in SN 1998bw, as obtained from our modelling in Sections
\ref{2002ap} and \ref{1998bw}; see also \citet{Mazzali01,Mazzali07b}]
and the degree of ionisation of oxygen
is rather high (roughly 10 $-$ 50$\%$; Figure \ref{pap4test4}) it is a reasonable
simplification to neglect all elements besides the iron-group elements
and oxygen when calculating the ionisation balance and to assume that all electrons are provided by oxygen and by iron, which is
assumed to be 100$\%$ singly ionised. Therefore
\begin{equation}
Y'_\mathrm{O[I]}n_\mathrm{O[I]} \sim R_\mathrm{O[I]}(n_\mathrm{O[II]}
+ n_\mathrm{Fe})\zeta n_\mathrm{O[II]}
\label{ionbal}
\end{equation}
which gives
\begin{equation}
\begin{split}
\label{nn}
\frac{n_\mathrm{O[I]}}{n_\mathrm{Tot}} & \sim 1
-\frac{n_\mathrm{Fe}}{n_\mathrm{Tot}} 
 - \frac{R_\mathrm{O[I]}n_\mathrm{Fe}\zeta +
  Y'_\mathrm{O[I]}}{2R_\mathrm{O[I]}n_\mathrm{Tot}\zeta}\\
& \times \Bigg(-1 + \sqrt{1 +
  \frac{4Y'_\mathrm{O[I]}R_\mathrm{O[I]}(n_\mathrm{Tot} - n_\mathrm{Fe})\zeta}{(R_\mathrm{O[I]}n_\mathrm{Fe}\zeta
    + Y'_\mathrm{O[I]})^2}}\Bigg)\\
& \sim 1 -\frac{n_\mathrm{Fe}}{n_\mathrm{Tot}} + \frac{R_\mathrm{O[I]}n_\mathrm{Fe}\zeta +
  Y'_\mathrm{O[I]}}{2R_\mathrm{O[I]}n_\mathrm{Tot}\zeta} \\
 & - \sqrt{\frac{Y'_\mathrm{O[I]}(n_\mathrm{Tot} -
     n_\mathrm{Fe})}{R_\mathrm{O[I]}n_\mathrm{Tot}^2\zeta}} \ ,\hspace{1cm} \frac{Y'_\mathrm{O[I]}(n_\mathrm{Tot} -
     n_\mathrm{Fe})}{R_\mathrm{O[I]}n_\mathrm{Tot}^2\zeta} \gg 1\\
& \sim 1 - \frac{n_\mathrm{Fe}}{n_\mathrm{Tot}}\\
& -
 \frac{Y(n_\mathrm{Tot}-n_\mathrm{Fe})}{(R_\mathrm{O[I]}n_\mathrm{Fe}\zeta
   + Y'_\mathrm{O[I]})n_\mathrm{Tot}} \ , \hspace{0.45cm} \frac{Y'_\mathrm{O[I]}(n_\mathrm{Tot} -
     n_\mathrm{Fe})}{R_\mathrm{O[I]}n_\mathrm{Tot}^2\zeta} \ll 1
\end{split}
\end{equation} 
In Figure \ref{pap4test4} this ratio is computed for a test model
described in Appendix A using clumping factors of 1, 10 and 100.
The total luminosity of any recombination line X of oxygen relative to
the total deposited luminosity is then given by
\begin{equation}
\begin{split}
\frac{L_\mathrm{X}}{L_\mathrm{Dep}} & =
Y'_\mathrm{O[I]}n_\mathrm{O[I]}VE_\mathrm{X}f_\mathrm{X} L_\mathrm{Dep}^{-1}\\
             & = (1 + {\cal R})\frac{E_\mathrm{X}f_\mathrm{X}}{W_\mathrm{O[I]}} \Big(\frac{n_\mathrm{O[I]}}{n_\mathrm{Tot}}\Big)\\
\label{Lx}
\end{split}
\end{equation}
where $E_X$ is the energy and $f_X$ is the normalised effective
recombination fraction of the line. Usually one can assume 4$Y'_\mathrm{O[I]}R_\mathrm{O[I]}n_\mathrm{Tot}\zeta
\gg (R_\mathrm{O[I]}n_\mathrm{Fe}\zeta + Y'_\mathrm{O[I]})^2 $
; see Appendix~A. $L_{Dep}$ \citep[e.g.][]{Axelrod80} and all
other variables (besides the temperature) can be calculated for any
ejecta model and
 Equation \ref{Lx} can be evaluated directly. The
 temperature can be determined by balancing heating and cooling taking into
 account several hundred emission lines (therefore an exact estimate of
 the temperature is difficult) and varies between
 8000~K and 2000~K at the epochs of interest (100  to 600 days). A
 sufficiently exact estimate can be obtained assuming a constant
 temperature of $\sim$ 4000~K (see Appendix~A).

The luminosity of any oxygen recombination line is directly
proportional to the normalised effective recombination fraction $f_\mathrm{X}$ and the energy of the
transition $E_\mathrm{X}$ [eV]. The
recombination line luminosity is influenced only weakly by the
clumping factor $\zeta$ (which changes the ionisation
fraction). The oxygen NGS recombination
lines will be weak in general [taking
$n_\mathrm{O[I]}$/$n_\mathrm{Tot} \sim$ 0.5, $f_{7774} \sim 0.4$, $E_{7774} \sim$
1.6~eV and $W_\mathrm{O[I]} \sim $ 75~eV the strongest NGS oxygen
recombination line (7774~\AA ) carries less than 1.0\% of the total
luminosity] and we will neglect all other NGS oxygen
recombination lines in the rest of this paper. They are calculated by
the nebular code, but in the nebular spectra
of SNe 1998bw and 2002ap on oxygen recombination lines other than 
7774~\AA\ can be identified. 

\subsection{Excitation of the O {\sc i} 7774~\AA\ line}
\label{oxex}

We will find in Sections \ref{2002ap} and \ref{1998bw} that the luminosity
provided by recombination is not sufficient to explain the
observations of the
7774~\AA\ line earlier than $\sim$ 250 days. Therefore, in this section we investigate other
excitation mechanisms which might operate simultaneously with recombination.

Ground-state excitation by thermal electrons can clearly be ruled out. All
transitions from n $\ge$ 3 to the ground state have energies larger than 9~eV,
which makes thermal excitation by electrons of temperatures $\sim 0.5$
eV ineffective. 

Non-thermal excitation by the same electrons causing ionisation
(which are produced by Compton-scattering of $\gamma$-rays emitted by
$^{56}$Co decay) is also too weak. The
ratio of the cross-sections for electron impact excitation and
ionisation (at the high electron energies which are of interest here) can
be approximated by \citep[e.g.][]{Rozsnyai80}
\begin{equation}
\frac{\sigma_\mathrm{nl \rightarrow n'l'}}{\sigma_\mathrm{nl}} \sim
1.66\frac{f_\mathrm{nl \rightarrow n'l'}E_\mathrm{nl}}{N_\mathrm{nl}E_\mathrm{nl \rightarrow n'l'}} 
\end{equation}
where E$_\mathrm{nl \rightarrow n'l'}$ is the transition energy,
E$_\mathrm{nl}$ the ionisation energy, $N_\mathrm{nl}$ is the number
of electrons in the shell n, l and f$_\mathrm{nl \rightarrow n'l'}$ the
oscillator strength of the transition. We compare hydrogen and helium
non-thermal excitation rates obtained by this approximation to values
obtained from more sophisticated calculations (Hachinger in prep.), solving an energy-balance equation
derived from the Spencer-Fano equation \citep{Xu91,Lucy91} and
find that the agreement is always better than 30\%, which suggests that this
approximation is also of acceptable accuracy for oxygen. For allowed
ground state transitions of neutral oxygen the
oscillator strengths are of the order of
10$^{-2}$ decreasing with increasing quantum number
\citep{Bell90}. The total excitation rate into triplet states is
$\sim$ 5\% of the ionisation rate only. The excitation rates of quinted
states by non-thermal (high-energetic) electrons is even lower, since the excitation
cross-sections of spin-forbidden lines strongly decrease with electron
energy, as compared to allowed transitions \citep[e.g.][]{Ralchenko08}. 

Therefore, non-thermal
excitation is not important compared to
recombination for populating excited levels. In addition, if the O {\sc i} 7774~\AA\ line was
excited by non-thermal electrons, one would expect the ratio of
the line to the total luminosity to be approximately constant at all epochs,
which is clearly not observed (this ratio in fact decreases considerably at
late epochs).

Although it is generally assumed that the SN is optically thin during the nebular
phase, some lines can still be optically thick (especially ground state
transitions). However, as we show here, the optical depth of the
7774 \AA\ transition is still high, even at epochs of $\sim$ 200 days
because the 3s($^5$S) level is populated by recombining electrons.

We  estimate the time-dependent optical depth of
the 7774~\AA\ line resulting from recombination taking into account
the effective recombination rates, resonance scattering of the
3s($^5$S) to 2p($^3$P) transition as well as the recycling of ground state recombination radiation. 

Based on Equations \ref{ionbal} and \ref{nn}, the number density of
the 3s($^5$S) state relative to the total density is
approximately given by
\begin{equation}
\begin{split}
\frac{n_{3s(^5S)}}{n_\mathrm{Tot}} & \sim  \frac{f_{3s(^5S)}R_\mathrm{O[I]}(n_\mathrm{Fe} +
  n_\mathrm{O[II]})\zeta n_\mathrm{O[II]}}{(A^\prime + C)n_\mathrm{Tot}} \\
         & \sim
\frac{f_{3s(^5S)}Y'_\mathrm{O[I]}\tau}{A}\Big(\frac{n_\mathrm{O[I]}}{n_\mathrm{Tot}}\Big)\ ,\qquad A \gg A^\prime \gg
C\\
& \sim
(1 +
{\cal R})\frac{f_{3s(^5S)}\tau_0L_\mathrm{Dep}}{AVW_\mathrm{O[I]}}\Big(\frac{n_\mathrm{O[I]}}{n_\mathrm{Tot}}\Big)^2\zeta\\
\label{n6}
\end{split}
\end{equation}
defining $A^\prime$ as the reduced (due to self-absorption) radiative rate
\begin{equation}
A^\prime = A\frac{1 - \exp(-\tau)}{\tau}\sim \frac{A}{\tau}\ , \qquad \tau \gg 1
\end{equation}
from the 3s($^5$S) to the ground state, where $\tau$ is the Sobolev optical
depth of this ground state transition
\begin{equation}
\tau  \sim \frac{\lambda^3_{1355}tg_\mathrm{3s(^5S)}A_{1355}n_\mathrm{O[I]}\zeta}{8\pi
  g_\mathrm{2p(^3P)}} \sim \tau_0\zeta n_\mathrm{O[I]}\ ,
\hspace{0.2cm} n_\mathrm{O[I]} \gg n_{3s(^5S)}\\
\end{equation}
with the ground state transition wavelength $\lambda_{1355}$ =
1355~\AA, $t$ the epoch in seconds, $V$ the volume in cm$^3$, $A_{1355}$ the
radiative rates from the 3s($^5$S) to the ground state
(note that the ground state is split into three j sub-levels and
appropriate weights have to be used; also the radiative rates of these
j sub-state transitions differ). The weights of the
upper and lower states are $g_\mathrm{3s(^5S),2p(^3P)}$; $f_{3s(^5S)}$
is the normalised effective recombination fraction into the
3s($^5$S) state and C is the de-excitation rate due to thermal electrons which is
much smaller than $A/\tau$. The 3s($^5$S) to 2p($^3$P) transition is
spin-forbidden but dipole-allowed and therefore collisional rates are
weaker than radiative rates by several orders in magnitude
\citep{Regemorter62} for typical SN nebular densities. For
$n_\mathrm{O}\zeta > 10^{9}$ cm$^{-3}$ this assumption breaks down since C and $\tau$ will increase with density and clumping factor.

Equation \ref{n6} can be used to calculate the Sobolev optical depth
of the 7774~\AA\ line
\begin{equation}
\begin{split}
\tau_{3s(^5S)} & \sim \frac{\lambda^3_{7774}tg_\mathrm{3p(^5P)} A_{7774}n_\mathrm{3s(^5S)}[\zeta]}{8\pi
  g_\mathrm{3s(^5S)}} \ , \hspace{0.4cm} n_\mathrm{3s(^5S)} \gg n_{3p(^5P)}\\
\label{tau6}
\end{split}
\end{equation}
with the transition wavelength $\lambda_{7774}$ = 7774~\AA , $t$ the epoch in seconds, $A_{7774}$ the
radiative rates from the 3p($^5$P) to the 3s($^5$S) state
(note that the 3p($^5$P) state is split into three j sub-levels and
appropriate weights have to be used) and $g_\mathrm{3p(^5P),3s(^5S)}$ the weights of the
upper and lower states.
Again we compare this estimate with numerical results from our nebular
code in Appendix A.

At this point it is important review the concept of clumping in
more detail. It is assumed that the ejecta are distributed into small ``blobs'' with 
size much smaller than the typical scale of the SN, covering the SN
volume homogeneously \citep[e.g.][]{Li92}. Therefore, although the optical depth is
increased locally in the emitting region, the global optical depth of
the SN remains constant. This means that, when scattering radiation
from remote regions of the SN, the opacity is not influenced directly by
clumping. On the other hand, when scattering radiation in the emission
region, the optical depth increases proportionally to the clumping factor. 

In addition, there is an increase of the ratio
$\frac{n_{3s(^5S)}}{n_\mathrm{Tot}}$ (see Equation \ref{n6}), which results from the increased
self-absorption of ground state transition radiation. Therefore,
even when scattering radiation from remote regions, clumping 
increases the optical depth and so it does influence the
7774~\AA\ resonance scattering of background radiation.

 The 3p($^5$P) state is separated from the 3s($^5$S)
state by $\sim$ 1.6~eV, therefore thermal electron excitation is
possible, but the population of the 3s($^5$S) state by thermal electron
excitation is too low. Therefore the state must be populated by
recombining electrons.

The thermal electron excitation coefficient
from 3s($^5$S) to 3p($^5$P) is given by
\begin{equation}
\begin{split}
C_\mathrm{3p(^5P)} & \sim 8.6 \times
10^{-6}\frac{n_\mathrm{e}\zeta}{T^{1/2}}\frac{\Omega_{7774}}{g_\mathrm{3s(^5S)}}\exp(-\frac{E_{7774}}{kT})\\
& \equiv C_0n_\mathrm{e}\zeta
\end{split}
\end{equation}
where all constants have the same meaning as in Equation \ref{tau6},
$E_{7774}$ is the energy corresponding to 7774~\AA\ and
$\Omega_{7774}$ = 25.1/40.2 \citep[at 5000/10000 K;][]{Bhatia95} is 
the effective collision strength of the O {\sc i}
7774~\AA\ transition. One can compare the
population of the 3p($^5$P) state due to
thermal electron collisional excitation and recombination respectively by
calculating their ratio
\begin{equation}
\begin{split}
\mathcal{R}_\mathrm{3p(^5P)} & = \frac{C_\mathrm{3p(^5P)}n_\mathrm{3s(^5S)}}{Y_\mathrm{O[I]}n_\mathrm{O[I]}f_\mathrm{3p(^5P)}}\\
& = (1 +
{\cal
  R})C_0\frac{f_\mathrm{3s(^5S)}}{f_\mathrm{3p(^5P)}}\frac{\tau_0}{A}(n_\mathrm{Tot} - n_\mathrm{O[I]})n_\mathrm{O[I]}\zeta^2
\end{split}
\end{equation}
The ratio of the O {\sc i} 7774~\AA\ luminosity induced by thermal
electron excitation compared to the total
deposited luminosity is then given by
\begin{equation}
\begin{split}
\frac{L_{3p(^5P)}}{L_{Dep}} & \sim (1 + {\cal
  R})^2\frac{C_0E_\mathrm{7774}f_\mathrm{3s(^5S)}\tau_0}{W_\mathrm{O[I]}A}(1
- \frac{n_\mathrm{O[I]}}{n_\mathrm{Tot}})n_\mathrm{O[I]}^2\zeta^2\\
& \sim 10^{-21}t\mathrm{[days]}(1
-
\frac{n_\mathrm{O[I]}}{n_\mathrm{Tot}})n_\mathrm{O[I]}^2\zeta^2, \hspace{0.5cm}
T = 5000~\mathrm{K}
\label{L7}
\end{split}
\end{equation}
This estimate becomes invalid for $n_\mathrm{O[I]}\zeta > 10^{9}$
cm$^{-3}$ since Equation \ref{n6} had been used for its
derivation. Equation \ref{L7} is compared to results from our nebular code in Appendix~A.
Whether collisional excitation of the O {\sc i} 7774~\AA\ line is
negligible depends mainly on the temperature, which decreases with time, on the density of neutral oxygen, which decreases with $t^{-3}$,
and on the clumping factor. Therefore, especially for early
epochs (high temperature, high density) and
for large clumping factors ($\zeta \gg 1$) the O {\sc i} 7774~\AA\ line
may be excited by thermal excitation of recombining electrons. Around
150 days the core (oxygen) density of CC-SNe is typically of order
$10^{7-8} $ cm$^{-3}$, which gives 
$\frac{L_{3p(^5P)}}{L_{Dep}} \sim (10^{-5} ~\mathrm{to}~
10^{-3})\zeta^2$ assuming a temperature of 5000 K.

From the above considerations it becomes clear that clumping influences
the recombination line strength only weakly, but increases the
effect of resonance scattering and thermal
excitation of the O {\sc i} 7774~\AA\ transition strongly. For both processes the 3s($^5$S)
population is provided by recombination and not by thermal electron
excitation from the ground level.

\subsection{Emission versus absorption line shapes}
\label{emabs}

\begin{figure} [h]
\begin{center}
\includegraphics[width=8.5cm, clip]{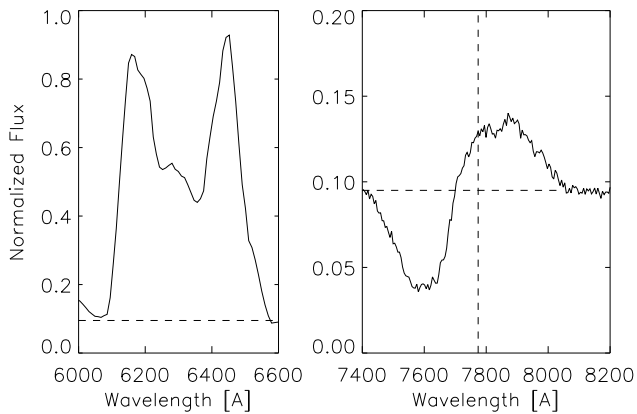}
\end{center}
\caption{Left side: flux of a synthetic [O {\sc i}] $\lambda\lambda$ 6300,6363 doublet for an asymmetric model (double-peaked shape). Right
  side: the 7774~\AA\ line opacity calculated from this asymmetric
  model was used to calculate the flux of the absorption line at 7774~\AA\ (dashed vertical line). The flux is normalised to an arbitrary constant. The
formation of the absorption line strongly depends on the assumed
background (indicated by
the dashed horizontal line; it not only depends on the absolute
background flux but also on the background rest-frame wavelength which
determines the
emission site) and therefore the absorption line shown would look
different for a different background radiation field. The asymmetry has some influence on the
absorption line, but the clear double-peaked shape of the [O {\sc i}]
$\lambda\lambda$ 6300,6363 doublet is not seen.}
\label{pap4abs}
\end{figure}

Since the [O {\sc i}] $\lambda\lambda$
6300,6363 doublet has often been used to probe the ejecta velocity and
the geometry of
CC-SNe cores (see Section \ref{intro}) it is interesting to study the profile of the [O
  {\sc i}] 7774~\AA\ line. If this line were caused by recombination
or thermal excitation alone, one would expect that the shapes of the 6300, 6363 and 7774
\AA\ lines should be at least approximately similar.

However, as we have shown above, O {\sc i} 7774~\AA\ might
result from line scattering
even at 200 days. At later times the line usually becomes
very weak and it is difficult to use it to probe the ejecta geometry. 
The profile of a scattering dominated line depends on the flux
distribution, so one cannot
expect that the shape of the 6300, 6363~\AA\ lines and the 7774~\AA\ line will
be even approximately similar. 

To demonstrate this, we use a multi-dimensional Montecarlo code. A
photon background is generated and resonance-scattered in an (asymmetric) oxygen
distribution (e.g. taken from a nebular model; the nebular code in its
current version
cannot simulate scattering processes). We calculate the [O {\sc i}] $\lambda\lambda$ 6300,6363
doublet for an asymmetric oxygen model and use the 7774~\AA\ line
opacity calculated in the nebular code to compute the absorption
of a broad-band background (3000 $-$ 10000~\AA ) emitted in the inner region of the
model SN. The formation of the absorption line is very sensitive to this
background and therefore our results (see Figure \ref{pap4abs})
illustrate just one of many possibilities. As expected, the
absorption line has quite a different shape than the emission line.

\subsection{Ionisation Rates for the nebular code}
\label{ion}
To improve the accuracy of the recombination line calculation in our
nebular code we updated the treatment of ionisation for all the ions treated. This
is necessary, since the formation of recombination lines is much
more sensitive to the ionisation rate than the formation of other
emission lines, and higher accuracy is needed. In the older version of
the nebular code
the ionisation rates were calculated using a simple analytical form.

To obtain more accurate ionisation rates we calculate the average `work per ion' $W$ \citep[see][]{Axelrod80}, which is the amount of energy lost by a non-thermal
electron (with energy $\gg$ 1~keV) to cause one
ionisation. The ionisation rate is then computed dividing the
total deposited luminosity by this energy. 

The value of $W$ can be calculated averaging the `work per ion' $W^*_\mathrm{i}$ of all individual
ions  over the total ionisation
cross sections. $W^*_\mathrm{i}$ can be obtained for any ion
integrating the ratio of the energy-dependent ionisation cross section
$\sigma_\mathrm{i}$ and the energy loss functions $L_\mathrm{i}$
\citep{Axelrod80}
\begin{equation}
W^*_\mathrm{i} = \frac{E_\mathrm{max}}{\int_{E_\mathrm{min}}^{E_\mathrm{max}}\frac{\sigma_i(E')}{L_\mathrm{i}(E')}dE'}
\end{equation}
 
Highly energetic electrons lose most of their
energy to ionisation and to the secondary electrons produced in the
ionisation, but also by non-thermal excitation. At low energies, losses
to the electron plasma become
important, however most of the primary energy is lost in
atomic collision processes. 

Ionisation cross sections and loss functions can be
obtained for any ion from the literature at low electron energies
\citep[e.g.][]{Lotz67,Lotz70,Lennon88}. At high electron energies they
can be computed by appropriate extrapolations of the low energy values \citep{Axelrod80}. 

The ionisation rate of any ion is then given by
\begin{equation}
Y_\mathrm{i} =
\frac{L_\mathrm{Dep}\sigma_\mathrm{i}}{N_\mathrm{Tot}W<\sigma>} \equiv
\frac{L_\mathrm{Dep}}{N_\mathrm{Tot}W_\mathrm{i}}
\end{equation}

The ionisation rates
mainly affect the recombination lines but also have some influence on all other
emission lines. This will be discussed in Section \ref{disc}.

\section{A shell model of SN 2002ap}
\label{2002ap}

SN 2002ap is classified as a broad-lined SN of Type Ic. It had an
ejected mass of $\sim$ 2.5  M$_\odot$ and a kinetic energy of $\sim$ 4
$\cdot\ 10^{51}$~ergs \citep{Mazzali07b}. The
distance and reddening to SN 2002ap are only known approximately. To be consistent with previous
work we use $\mu$ = 29.50 mag and $E(B - V)$ = 0.09~mag as done by
\citet{Mazzali02,Yoshii03,Mazzali07b}. It is not clear
whether SN 2002ap was a spherical symmetric event or not. Recently,
\citet{Maurer09} have shown that an asymmetry might be observable in all
broad-lined CC-SNe, however this point is not clear yet. The
spectra of SN 2002ap used in this work are those used by
\citet{Mazzali07b}.

\citet{Mazzali07b} found appropriate one-zone and shell models of the SN 2002ap nebular ejecta. We quickly summarise the
main findings for the shell models here. Using a filling factor of
0.1 ($\zeta$ = 10) in the $^{56}$Ni rich regions, a total ejecta mass of $\sim$ 2.5
M$_\odot$ was found, containing roughly 0.11 M$_\odot$ $^{56}$Ni and
1.3 M$_\odot$ of oxygen.

In contrast to this previous work, which aimed at modelling each
spectrum individually, here we try to find one single $^{56}$Ni/O model which can produce the
time evolution of the different spectra at all observed epochs
consistently. Special attention is paid to reproducing the
exact shape of the [O {\sc i}] $\lambda\lambda$ 6300,6363 line
profile, which is a tracer of
the distribution of oxygen, the most abundant element of the nebula.

\begin{figure} 
\begin{center}
\includegraphics[width=8.5cm, clip]{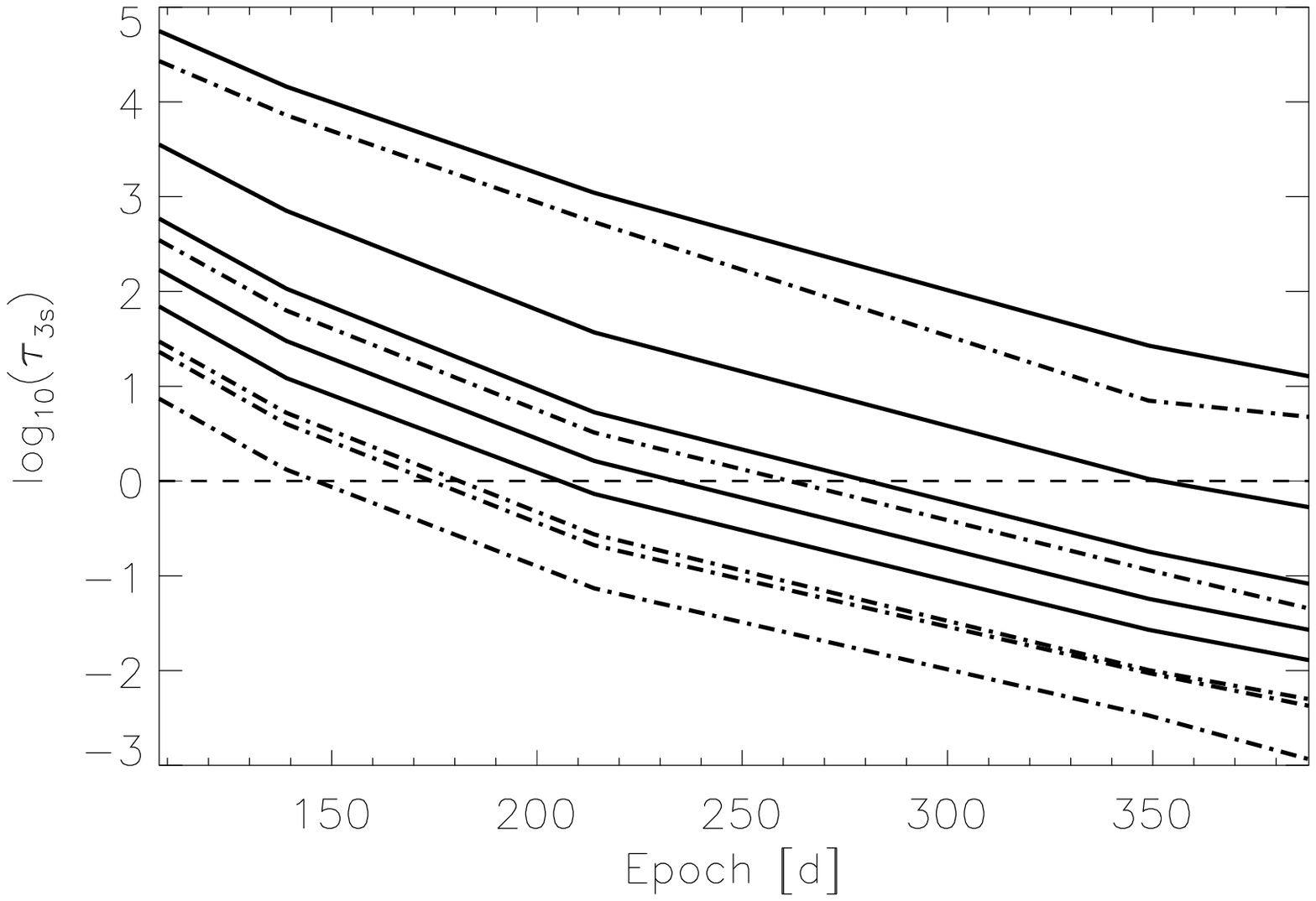}
\end{center}
\caption{SN 2002ap, models for a clumping factor $\zeta$ = 5. Left side: normalised nebular spectra (from top to bottom: 129, 163,
  192, 229, 245, 253, 281, 343 and 394 days after explosion) are shown
  in black with synthetic spectra on top in red. Right side: the
region between 7600~\AA\ and 8200~\AA\ is enlarged to enable a
comparison of observed (black) and synthetic (red) 7774~\AA\ flux. At
epochs of 129 to 229 days, the simulated flux is too low to match the
observations. Beginning at 245 days, considering a constant off set
caused by background-flux which is not reproduced by the
synthetic spectrum, the synthetic 7774~\AA\ line starts to become consistent with
the observations.}
\label{pap42002apc}
\end{figure}

\begin{figure} 
\begin{center}
\includegraphics[width=8.5cm, clip]{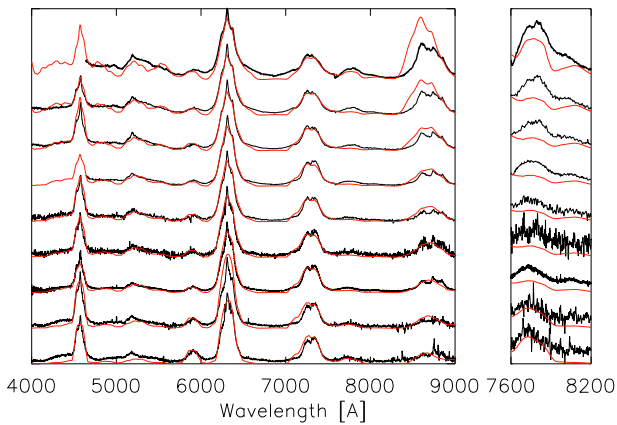}
\end{center}
\caption{SN 2002ap, models for a clumping factor $\zeta$ = 25. Left side: normalised nebular spectra (from top to bottom: 129, 163,
  192, 229, 245, 253, 281, 343 and 394 days after explosion) are shown
  in black with synthetic spectra in red. Right side: the
region between 7600~\AA\ and 8200~\AA\ is enlarged to enable a
comparison of observed (black) and synthetic (red) 7774~\AA\ flux. At
epochs of 129 and 163 days, the simulated flux matches the
observations at 7774~\AA\ much better than for low clumping factors but in general the fit to the forbidden line emission is
worse (e.g. the [O {\sc i}] 5577~\AA\ line and the Ca {\sc ii}
IR-triplet line at $\sim$ 8500~\AA\ become
too strong at early epochs). Around 200 days line scattering seems
necessary to explain the observed 7774~\AA\ flux.}
\label{pap42002apf}
\end{figure}

Using the same distance modulus and reddening as \citet{Mazzali07b}
and 
clumping factors of 5 and 25 [we had to use
a clumping factor of 1 for the innermost shell of the
$\zeta = 25$ model to avoid the formation of sharp high density lines
which are not observed; \citet{Mazzali07b} had used a clumping
factor of 10; this will be discussed in Section \ref{disc}]
we obtain models similar to the ones given in \citet{Mazzali07b}
giving reasonable agreement with the observations at all epochs (see
Figures \ref{pap42002apc} and \ref{pap42002apf}). The forbidden oxygen lines are
reproduced at all epochs by one and the same oxygen distribution, which provides evidence for the reliability of this
model. 

Our $^{56}$Ni zone with a total mass of $\sim$ 1.2 M$_\odot$ extends
out to 8000~km/s, containing $\sim$ 0.07
M$_\odot$ of $^{56}$Ni \citep[in agreement with][]{Mazzali02} and
$\sim$ 0.7 M$_\odot$ of O. More
mass is located at higher velocities, but the
nebular modelling becomes inaccurate for the outer regions. The exact values depend on the
clumping factor as well as on the $^{56}$Ni
distribution, which is not known.

The O {\sc i} 7774~\AA\ line is excited by recombination and thermal
electron excitation in our nebular
modelling (no line scattering) and is too weak to
explain the observations at 129 and 163 days using low clumping
factors ($\zeta \sim 5$). The $\zeta = 25$ model can reproduce the
observations of the O {\sc i} 7774~\AA\ line at 129 and 163 days
better than the $\zeta = 5$ model. This results from thermal electron
scattering of 3s($^5$S) electrons, which are provided by
recombination. In general the $\zeta = 5$ model reproduces the
formation of the
forbidden lines better than the $\zeta = 25$ model at early epochs,
since high density lines, which are not observed, form in the $\zeta =
25$ model.

At days 192 and 229 both models produce too little flux at
7774~\AA , which is probably influenced by line scattering at these
epochs (the optical depth of the inner shells of the O {\sc i} 7774~\AA\ line is high for
both models at these epochs). The scattering process is not simulated
and therefore the synthetic flux is too low.
  
At later epochs the O {\sc i} 7774~\AA\ line is consistently reproduced by recombination alone in
both models when taking
into account the uncertainty due to the background around 7774~\AA , which is not reproduced by
the nebular code. 

\begin{figure} 
\begin{center}
\includegraphics[width=8.5cm, clip]{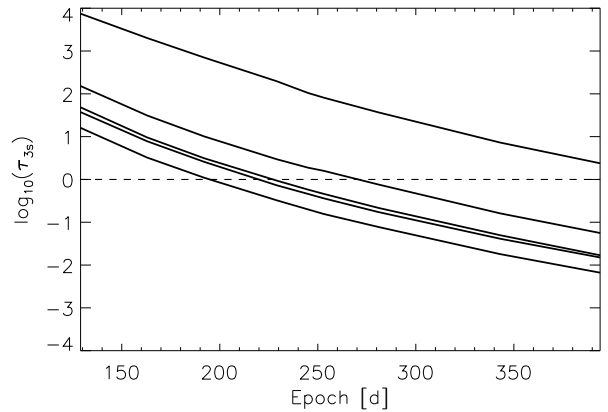}
\end{center}
\caption{SN 2002ap, models for a clumping factor $\zeta$ = 5. The logarithm of the 7774~\AA\ line optical depth in our
  model for the inner shells (from top to bottom:
0 $-$ 1000~km/s, 1000 $-$ 2000~km/s, 3000 $-$~4000 km/s, 4000 $-$ 5000~km/s
and 2000 $-$ 3000~km/s). The 2000 $-$ 3000~km/s shells contains very little
oxygen in order to reproduce the narrow peak on top of the [O {\sc i}]
$\lambda\lambda$ 6300,6363 doublet lines and therefore it has low optical
depth. Apart from the innermost shell, the optical depth drops below one around 220 days. This optical depth is calculated for
scattering of remote emission radiation, which has a weaker dependence
on the clumping factor than for local scattering.}
\label{pap42002apctau}
\end{figure}

The optical depth of the 7774~\AA\ line is shown for all inner shells
(five shells between 0 and 5000~km/s) of our $\zeta = 5$ model in
Figure \ref{pap42002apctau}. It does not fall below one before 210 days,
indicating that there can be strong line scattering up to
$\sim$ 200 days depending on clumping (for the $\zeta = 25$ model the
optical depth is higher and the O {\sc i} 7774~\AA\ line becomes
optically thin later).  

For clumping factors below 5 neither line scattering nor thermal
excitation was strong enough to allow sufficient formation of the O {\sc i}
7774~\AA\ line before $\sim$ 200 days.

\section{A 2D model of SN 1998bw}
\label{1998bw}

SN 1998bw is classified as a broad-lined SN of type
Ic \citep[ejected mass $\sim$ 14 M$_\odot$ and kinetic energy
  $\sim$ 6 $\cdot\ 10^{52}$ ergs;][]{Nakamura00}. We use a distance modulus of $\mu$ = 32.89 and an
extinction of A$_V$ = 0.2 mag as given by \citet{Patat01}. SN 1998bw was
accompanied by the low energy, long-duration GRB 980425. Due to this
GRB-SN connection there are speculations that SN 1998bw was a
highly aspherical event. These speculations are supported by
polarisation measurements \citep{Kay98,Iwamoto98,Patat01}, which where interpreted as a SN axis
ratio of 2:1 \citep{Hoflich99}. Some indication for core ejecta asymmetry had
also been found in the nebular spectra of SN 1998bw
\citep{Mazzali01}. The spectra of SN 1998bw used in this work are the
same as used by \citet{Mazzali01} originally presented by \citet{Patat01}.

The nebular phase of SN 1998bw has been well studied by means of nebular modelling \citep{Mazzali01,Maeda06}. While \citet{Mazzali01} attempted to model individual nebular
spectra by one-zone models (concluding that this is insufficient),
\citet{Maeda06} computed two-dimensional synthetic nebular
spectra for models obtained from hydrodynamical simulations, finding good
agreement with spectra of individual epochs. However no model has so far described the time sequence of nebular spectra
of SN 1998bw consistently. 

In this section we try to find a single two-dimensional model
which is consistent with the full time evolution from 108 to 388 days
after explosion. While successful models for individual epochs can be found even
in one dimensional modelling, it seems impossible to model
the time sequence of all spectra with a single one-dimensional model (in contrast to SN
2002ap, where this approach works well).

Since SN 1998bw is quite massive compared to other
CC-SNe (like SN 2002ap for example), the transition to the nebular
phase occurs at rather late times, although the ejecta expand at
high velocities. The spectra at 108 and 139 days
after explosion are not strictly nebular and therefore large deviations
between the model and the observations are to be expected when trying to reproduce these spectra
(especially at 108 days) with our nebular code.

When considering two dimensions one is immediately confronted with the
problem that the parameter space is much larger [angular
distribution of elements, observer inclination]. In addition, the
computation time to obtain one synthetic spectrum increases dramatically. Since hundreds of spectra have to be computed to
obtain a good model, this technical problem can make modelling
unfeasible. Therefore some assumptions about the SN geometry
have to be introduced.

There are several good arguments to assume that SN 1998bw might
consist of some kind of two-dimensional 'jet + disc' structure, as
described by for example by \citet{Maeda06} (see Section
\ref{intro}). Furthermore, since the SN was accompanied by a GRB, it seems likely that the
observer inclination is not too far from polar, although this is
uncertain since the opening angle of GRB 980425 is not known.

Therefore we work with a parametrised two-dimensional model which
consists of a polar zone (0$^\circ$ to 45$^\circ$) and an equatorial zone
(45$^\circ$ to 90$^\circ$). This simplification will introduce some error,
since most likely it does not represent physical
reality. However, it seems sufficiently exact to obtain an
acceptable fit at all epochs with a single $^{56}$Ni/O model. A model with more degrees
of freedom is highly degenerate anyway, since the
information that can be extracted from a series of spectra F($\nu$,t) is limited. 

Because of this simplification the reliability of our model is in
question, but this kind of uncertainty is inherent to all
multi-dimensional modelling. We performed tests for viewing angles of
0$^\circ$, 15$^\circ$ and 30$^\circ$. Using the smaller viewing angles
the iron lines which are mainly produced in the jet-like structure become very
narrow, which is not observed. Therefore we decided to use a viewing
angle of 30$^\circ$. However this might be a consequence of our
simplified geometry and therefore it is not possible to obtain
valuable information about the observers inclination from our
modelling approach.

Our model is roughly
consistent with previous findings \citep{Maeda06}. The polar zone
contains $\sim$ 0.24 M$_\odot$ of $^{56}$Ni at velocities below 12000
km/s. The equatorial zone also contains  $\sim$ 0.24 M$_\odot$ of
$^{56}$Ni (note that the equatorial zone has more than twice the
surface area of the polar zone), but located at velocities below $\sim$ 8000
km/s. The total mass below 12000~km/s is estimated to be $\sim$ 2.7 M$_\odot$
containing a total oxygen mass (below 12000~km/s) of $\sim$
2 M$_\odot$. There is a lot of material at higher velocities,
but the nebular modelling becomes inaccurate for the outer regions
because the density is too low.

\begin{figure} 
\begin{center}
\includegraphics[width=8.5cm, clip]{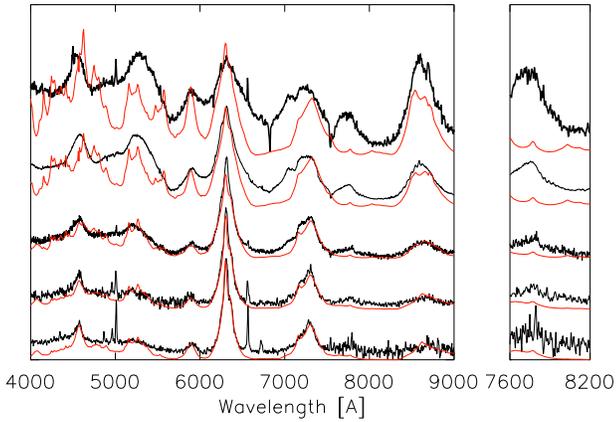}
\end{center}
\caption{SN 1998bw, case of no clumping ($\zeta = 1$). Left side: normalised nebular spectra of SN 1998bw (from top to bottom: 108, 139,
  214, 349 and 388 days after explosion) are shown
  in black with the synthetic spectra in red. On the right side the
region between 7600~\AA\ and 8200~\AA\ is enlarged to enable a
comparison of observed (black) and synthetic (red) 7774~\AA\ flux. The two earliest spectra (108, 139 days) of SN 1999bw are
clearly not nebular and the agreement between observations and simulations
is expected to be poor. However, the evolution of total and oxygen
6300~\AA\ flux as well as the shape of the [O {\sc i}] $\lambda\lambda$ 6300,6363
doublet line are reproduced at all epochs much better than by a one-dimensional model.  At
epochs of 108 to 349 days, the simulated flux is too low to match the
observations of the 7774~\AA\ line. At 388 days, considering the constant offset
caused by some background flux which is not reproduced by the
synthetic spectrum, the synthetic 7774~\AA\ line  seems to be
consistent with the observations. Unfortunately the noise level is high
and a detailed comparison is not possible.}
\label{pap41998bw35}
\end{figure}

\begin{figure} 
\begin{center}
\includegraphics[width=8.5cm, clip]{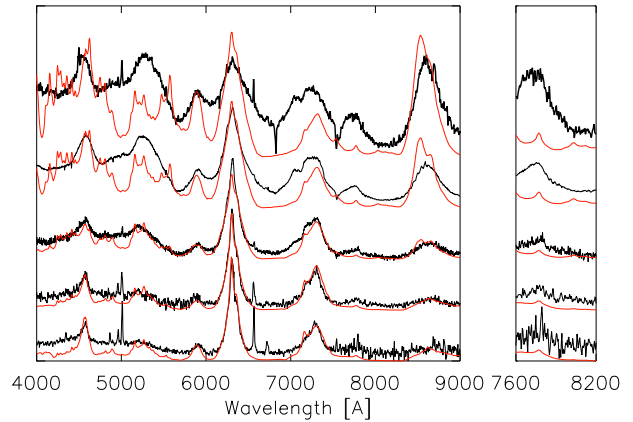}
\end{center}
\caption{SN 1998bw, case of clumping ($\zeta = 5$). Left side: normalised nebular spectra (from top to bottom: 108, 139,
  214, 349 and 388 days after explosion) are shown
  in black with the synthetic spectra in red. Right side: the
region between 7600~\AA\ and 8200~\AA\ is enlarged to enable a
comparison of observed (black) and synthetic (red) 7774~\AA\ flux. The two earliest spectra (108, 139 days) of SN 1999bw are
clearly not nebular and the agreement between observations and simulations
is expected to be poor. However, the evolution of total and oxygen
6300~\AA\ flux as well as the shape of the [O {\sc i}] $\lambda\lambda$ 6300,6363
doublet line are reproduced at all epochs much better than by a one-dimensional model.  At
epochs of 108 to 349 days, the simulated flux is too low to match the
observations of the 7774~\AA\ line. At 388 days, considering the constant offset
caused by some background flux which is not reproduced by the
synthetic spectrum, the synthetic 7774~\AA\ line  seems to be
consistent with the observations. Unfortunately the noise level is rather high
and a detailed comparison is not possible.}
\label{pap41998bwc}
\end{figure}

Again we find that the O {\sc i} 7774~\AA\ line is not reproduced as
due to
recombination by our model with clumping $\zeta = 1$ and $\zeta = 5$.
\citet{Mazzali01} and \citet{Maeda06} had used a clumping factor of 10. This will
be discussed in Section \ref{disc}] for the
spectra at 108, 139 and 214 days (see Figures \ref{pap41998bw35} and
\ref{pap41998bwc}).
At 349 and 388 days we can reproduce the O {\sc i} 7774~\AA\ line 
consistently with the observations. There is some background around 7774~\AA\ which is
not reproduced by the nebular code sufficiently exactly. Shifting the oxygen recombination
line upwards by the possible background, the agreement seems
reasonable for both clumping factors. However the large background-line ratio makes an exact
comparison impossible. 

We tried to explain the formation of O {\sc
  i} 7774~\AA\ at the earlier epochs (108, 139 days) by increasing the
clumping factor, which in turn increases the thermal electron excitation from the
3s($^5$S) to the 3p($^5$P) level. However, this does not seem to
work. Both [O {\sc i}] 5577~\AA\ and the Ca {\sc ii} IR-triplet
become very strong at clumping factors $> 10$, while the
synthetic flux around 7774~\AA\ increases but seems still too weak to
explain the observations. Owing to
the high density of SN 1998bw the nebular approach might be not suitable for
such early epochs (the strong
continuum which is observed is not reproduced). In addition there are
uncertainties on the atomic data (especially the collision strengths),
which may influence our conclusion.
 
\begin{figure} 
\begin{center}
\includegraphics[width=8.5cm, clip]{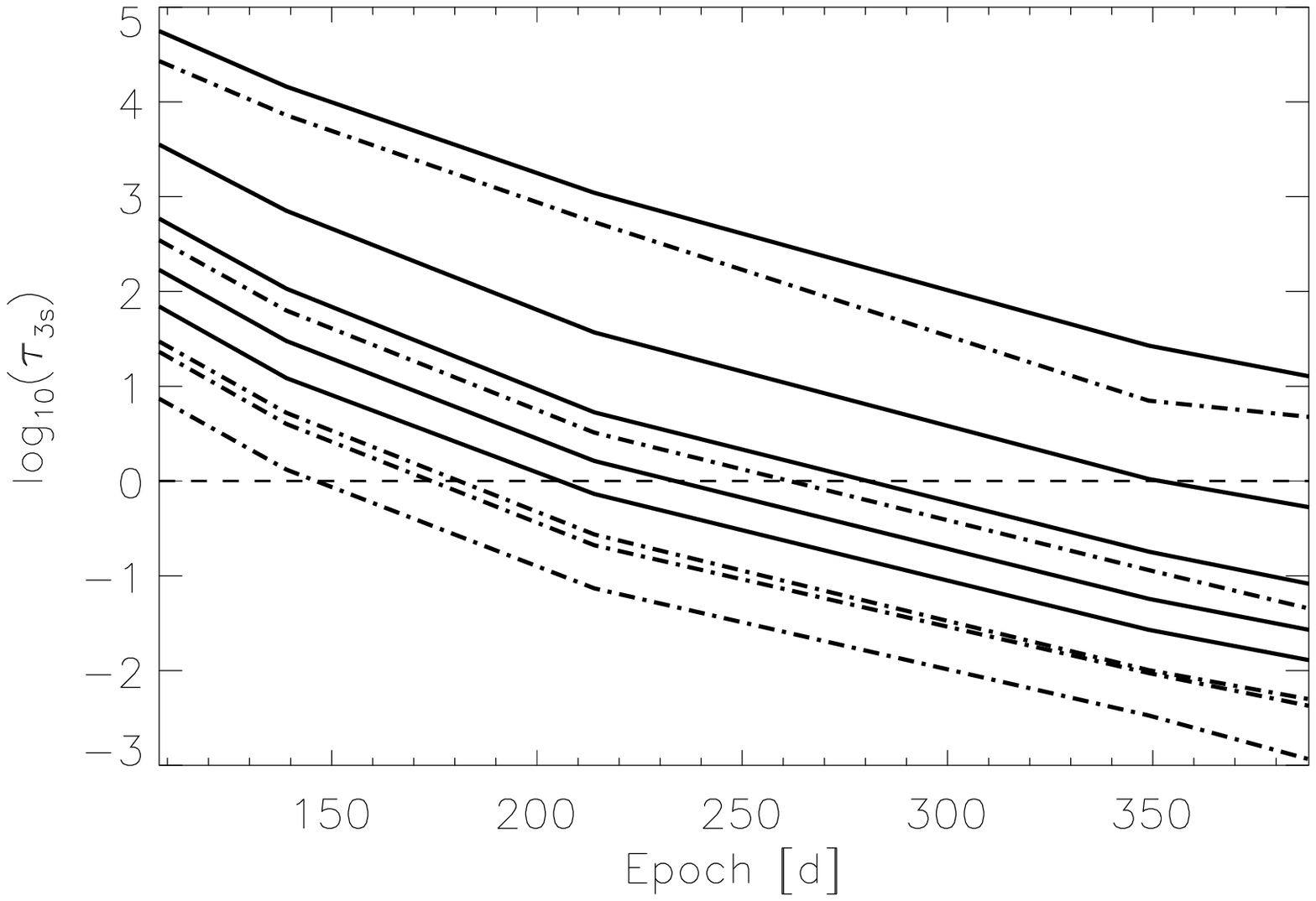}
\end{center}
\caption{SN 1998bw, case of no clumping ($\zeta = 1$). The logarithm of the 7774~\AA\ line optical depth in our
  model for the inner shells (from top to bottom:
0 $-$ 1000~km/s, 1000 $-$ 2000~km/s, 2000 $-$ 3000~km/s, 3000 $-$ 4000~km/s
and 4000 $-$ 5000~km/s). Solid lines show the equatorial region, dotted
lines the polar region. Apart from the innermost shell, the optical
depth drops below one at $\sim$ 210 days. This optical depth is calculated for
scattering remote emission radiation, which has a weaker dependence
on the clumping factor than for local scattering.}
\label{pap41998bw35tau}
\end{figure}

\begin{figure} 
\begin{center}
\includegraphics[width=8.5cm, clip]{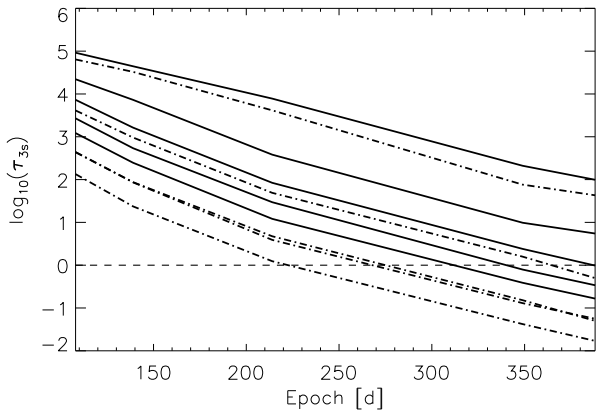}
\end{center}
\caption{SN 1998bw, case of clumping ($\zeta = 5$). The logarithm of the 7774~\AA\ line optical depth of our
  ejecta model for the inner shells (from top to bottom:
0 - 1000~km/s, 1000 - 2000~km/s, 2000 - 3000~km/s, 3000 - 4000~km/s
and 4000 - 5000~km/s). Solid lines show the equatorial region, dotted
lines the polar region. Apart from the innermost shell, the optical
depth drops below one at $\sim$ 310 days. This optical depth is calculated for
scattering remote emission radiation, which has a weaker dependence
on the clumping factor than for local scattering.}
\label{pap41998bwctau}
\end{figure}

 In Figures \ref{pap41998bw35tau} and
\ref{pap41998bwctau} we show the opacity of the 7774~\AA\ line for the
five innermost shells for clumping factors of $\zeta = 1$ and $\zeta
= 5$. There will be 7774~\AA\ line absorption up to $\sim 200 - 300$
days depending on clumping. 

Again, at least moderate clumping ($\zeta \sim 5$) seems necessary in order to provide
enough O {\sc i} 7774~\AA\ line opacity to reproduce the observations
at 214 days.

\section{Discussion}
\label{disc}
In Section \ref{effrec} we have obtained normalised effective
recombination fractions using available atomic data and computing
missing atomic data. The agreement with atomic data
available in the literature is good. Effective recombination
rates had been given before by \citet{Julienne74} at a temperature of
1160~K. These results are reproduced well. We extend the temperature
range up to 10000~K, which should be an upper limit for temperatures of
SN nebular ejecta between 100 and 600 days. Typical uncertainties in the
atomic data seem to have small influence. 

In Sections \ref{reclf} and \ref{oxex} we have obtained estimates for the
formation of oxygen recombination lines and line absorption by the
3s($^5$S) state of neutral oxygen. We have shown that the luminosity of recombination
lines is weak at any epoch. The influence of clumping on the
recombination line is weak as well. Clumping will increase the recombination
rate, but will decrease the number of O {\sc ii} ions at the same
time. The total number of recombinations, which is equivalent to the
total number of ionisations, will increase only slightly. Therefore clumping offers no direct way for
increasing O {\sc i} 7774~\AA\ emission considerably.

We have shown that there is line scattering of 7774~\AA\ photons even at very
late epochs. This is possible because of the large number of ground state
O {\sc i} atoms, which resonantly scatter the UV radiation
emerging from 3s to 2p transitions and slow down the
de-population of the 3s state. Combined with
recombination this leads to a high
population of the 3s level, which in turn can resonantly scatter
7774~\AA\ radiation. We have shown that, in contrast to the direct recombination
line, the scattering line is very sensitive to clumping of the
ejecta.

In addition, for high clumping factors, there may be thermal
electron collisional excitation of the O {\sc i}
7774~\AA\ line at early epochs. However, it is in question whether these high clumping
factors are realistic since high density
lines, which are not observed, may form.
Uncertainties in the atomic data, especially the collision
strengths, may influence our results. 

The considerations regarding the O {\sc i} 7774~\AA\ line are
also in principle
valid for the [O {\sc i}] 8446~\AA\ line [3s($^3$S) to 2p($^3$P)] with
the difference that the effective recombination rate into the 3s
triplet state is smaller than into the quintet
state. The optical depth of the 8446~\AA\ line is therefore expected
to be lower, which is still sufficient to cause
line scattering at very early epochs. However, the 8446~\AA\ line is blended with the blue wing of the [Ca
  {\sc ii}] IR-triplet, which can explain why no [O {\sc i}] 8446~\AA\ line is identified in SNe 1998bw and 2002ap.

We have shown that there is likely to be no other effective excitation mechanism
than 'recombination + thermal excitation' or 'recombination + absorption' for O {\sc i} n $\ge$ 3 levels. Thermal
electron excitation of n $\ge$ 3 levels from the ground level is ineffective at typical nebular
temperatures. Non-thermal electron excitation can also be ruled
out. First, it is too weak and second, it would produce a temporal
behaviour of the O {\sc i} 7774~\AA\ line which is not observed. As detailed modelling showed (Sections \ref{2002ap}  and
\ref{1998bw}) absorption might be too weak to explain the
observations as long as there is no clumping of the
ejecta ($\zeta \sim 5$). To enable thermal electron excitation even larger clumping
factors seem necessary.

The concept of clumping has often been used in stellar wind and SN
radiation physics \citep[e.g.][]{Li92,Li93,Mazzali01,Maeda06,Mazzali07b}
without quantitative understanding of the physics behind ejecta
clumping, therefore playing the role of a fitting factor.
\citet{Li92} have modelled the ratio of the [O {\sc i}] $\lambda\lambda$ 6300, 6363
doublet lines and \citet{Li93} the Ca {\sc ii} emission of SN 1987A using a clumping factor of $\sim$
10. \citet{Mazzali01,Maeda06,Mazzali07b} have modelled the nebular
spectra of SNe 1998bw and 2002ap finding that better fits to
observations are obtained using clumping factors of $\sim$ 10 than
when using no clumping. Our results are similar, but show some
difference. We obtain good fits to all
forbidden lines using clumping factors of 1 and 5 as well. The
abundance of elements will change slightly depending on the value chosen. We even found a
slight decrease in the quality of the fit when going from $\zeta = 5$
to $\zeta = 10$ and therefore we used the lower value. This difference
probably results mainly from our more accurate treatment of ionisation
which increases the ionisation rate of carbon and oxygen considerably.
This means
that the electron density is increased, which can be mimiced by 
increasing the clumping factor and using lower ionisation rates. It might
also be partially attributed to the fact that there is no strict criterion for the
quality of a fit for nebular spectra. 

Still, the constraints we derived for the clumping factor of SNe
1998bw and 2002ap from the O {\sc i} 7774~\AA\ scattering line compare well to previous findings. We found that a
clumping factor of $\sim$ 5 seems necessary to provide sufficient
opacity at 7774~\AA.
Clumping factors of $\zeta > 10$ make it possible to model
the O {\sc i} 7774~\AA\ emission at early epochs without line
scattering,
but may cause the formation of high density emission
lines at early epochs (100 $-$ 150 days). One example is the [O {\sc i}]
5577~\AA\ singlet line, which is observed to be weak. Since there are
uncertainties on the atomic data it is not clear whether
thermal electron excitation is important and whether the O {\sc i}
7774~\AA\ line at early epochs is dominated by line scattering or thermal electron
excitation of recombining electrons.

At later epochs, between 150 and 250 days, the line is
most likely a combination of line scattering and recombination
radiation, since thermal electron
excitation becomes too weak for any reasonable value of clumping. Around
$\sim$ 250 days the O {\sc i} 7774~\AA\ line seems to become a true
recombination line where the flux is mainly provided by the electrons
cascading into the 3p($^5$P) level.

In Section \ref{emabs} we have studied the expected profile of the [O
  {\sc i}] 7774~\AA\ line in light of our previous findings. As
expected, the absorption and emission line shapes caused by a certain
ejecta distribution can
show important differences. Further, the profile of the
absorption line depends on the background radiation field and one should be
careful when using the shape of a scattering line for any kind of
argumentation about ejecta geometry as long as it not clear how this
line is formed in detail.

In Sections \ref{2002ap} and \ref{1998bw} we obtained core ejecta
models for SNe 1998bw and 2002ap. The primary goal of this modelling
was not to re-derive ejecta properties, but to obtain reliable models
for studying oxygen recombination consistently with other line
formation. 

As a by-product we obtained
estimates for the $^{56}$Ni and oxygen masses of the cores of these
SNe. We find that SN 2002ap can be described very well by a
one-dimensional shell model (as already found before), which does not necessarily mean that
there is no asymmetry. Our $^{56}$Ni ($\sim$ 0.07 M$_\odot$) and
oxygen mass estimates seem
consistent with previous work. For SN 1998bw one-dimensional modelling
seems to be  insufficient to obtain an acceptable fit at all epochs using
a single
model. An acceptable fit is obtained with a 
two-dimensional 'jet + disc' model (this had been found before). The fit could certainly be improved by
increasing the degree of freedom of the model, but this would become
extremely time-consuming. The model found in this work reproduces the
observations at all epochs with increasing accuracy at later times,
which is expected. The estimate of the $^{56}$Ni ($\sim$ 0.48
M$_\odot$) and oxygen mass are consistent with previous results.

Using these ejecta distributions we were able to reproduce the formation
of O {\sc i} 7774~\AA\ at late epochs. There is enough
line opacity at 7774~\AA\ at early and intermediate nebular epochs to
produce a strong scattering line. Since the background radiation
field is important for the calculation of this absorption line, we were
not able to show that our ejecta distribution reproduces the
observations at early epochs exactly, however this seems likely. If
there is strong clumping the O {\sc i} 7774~\AA\ line may be additionally
excited by thermal electrons. The oxygen distribution inferred from
forbidden line observations allows the reproduction of the allowed
oxygen lines at late epochs. Since different
physics are involved, this provides a test for the consistency of the nebular
modelling approach, which is passed for SNe 1998bw and 2002ap. 
    
Several other ions may produce absorption lines at wavelengths
between 4000 and 10000~\AA\ (the part of the spectrum that is usually
observed). Since the abundance of these elements is much lower than
that of oxygen, there will be no excited states with
sufficient population to cause significant line scattering (an
exception might be the Ca {\sc ii} IR-triplet). Allowed
ground state transitions with energies between $\sim$ 1 and 3~eV are
interesting since their optical depth can be high owing to the high
radiative rates of allowed lines and to the possibly
sufficiently high ground state populations of
these low abundance elements. The elements
of interest are Na {\sc i}, Mg {\sc i} and Ca {\sc ii}. These three
elements have low quantum-level transitions
with wavelengths of $\sim$ 5890~\AA\ (Na {\sc i}) , 4570~\AA\ (Mg {\sc
  i}) and 3950~\AA\ , 7300~\AA\ and  8500~\AA\ (Ca {\sc ii}), which can be present
in sufficient amounts in SN nebulae. Iron group element lines,
especially Fe {\sc ii}, may be optically thick as well. However, since
these ions are complex we do not treat them explicitly here.

If 0.01~M$_\odot$ of Na {\sc i} were to be distributed homogeneously
within 5000~km/s, the optical depth of the Na {\sc i} 5890~\AA\ line
would be
$\sim$ 6 $\times 10^5$ at 200 days (this result is obtained by
balancing thermal-electron excitation with radiative and collisional
de-excitation analogously to Section \ref{reclf}), scaling linearly with Na {\sc I} mass and
with epoch t$^{-2}$.

If 0.01~M$_\odot$ of Mg {\sc i} were to be distributed homogeneously within
5000~km/s, the optical depth of the Mg {\sc i}] 4570~\AA\ line would be
$\sim$ 1 at 200 days, scaling linearly with Mg {\sc I} mass and
with epoch t$^{-2}$.

Three Ca {\sc ii} lines might be seen in
absorption even at late epochs ($\sim$ 3950~\AA , 7300~\AA\ and the IR-triplet 8500~\AA ). To estimate their strength, it is important to
know the relative population of the Ca {\sc ii} 3d to 4s states, which can be
approximated by
\begin{equation}
\frac{n_\mathrm{3d}}{n_\mathrm{4s}} \sim \frac{C_\mathrm{4s
    \rightarrow 3d}}{A^\prime_\mathrm{3d \rightarrow 4s} + C_\mathrm{3d
    \rightarrow 4s}} 
\end{equation}
where $C_\mathrm{4s \rightarrow 3d}$ and  $C_\mathrm{3d \rightarrow
  4s}$ are the collisional rates for the 4s to 3d transition and $A^\prime$ is
the radiative rate from 3d to 4s, reduced due to resonance scattering.

If 0.01~M$_\odot$ of Ca {\sc ii} were to be distributed homogeneously within
5000~km/s, the optical depth of the Ca {\sc ii} 3950~\AA\ and [Ca {\sc
    ii}] 7300~\AA\ lines would be
$\sim$ 2 $\times$ 10$^5$ and $\sim$ 0.02 respectively, at 200 days,
scaling linearly with the Ca {\sc ii} mass and
with epoch t$^{-2}$. The optical depth of the Ca {\sc ii} IR-triplet depends
on the population of the 3d state. Using the 7300~\AA\ optical depth
estimated above,
the optical depth of the Ca {\sc ii} 8500~\AA\ transition is $\sim$
430 ($n_\mathrm{e} = 10^7$ and T = 5000~K) at 200 days, with a stronger dependence on Ca {\sc ii} mass and epoch than
the optical depths at 3950~\AA\ and 8500~\AA\ \citep[see also][for a detailed discussion of Ca {\sc ii} line formation]{Li93}
. 

The exact mass of Na {\sc i}, Mg {\sc i} and Ca {\sc ii} is not well known, since estimates
of their masses are not very accurate. However, typical masses in CC-SNe
cores for Na, Mg and Ca might be of the order of 0.01 M$_\odot$.

Therefore, even at late times ($>$ 200 days) there is strong line scattering
at wavelengths between $\sim$ 4000 and 10000~\AA . The Ca {\sc ii}
3950~\AA , Na {\sc i} 5890~\AA\ and O {\sc i} 7774~\AA\ lines cause
strong scattering in any reasonable scenario. The [Mg {\sc i}]
4570~\AA , O {\sc i} 8446~\AA\ and Ca {\sc ii} 8500~\AA\  lines may have sufficient
optical depth for line scattering, depending on epoch, degree of
ionisation, total mass and clumping.   

\section{Summary and Conclusion}
\label{sum}
We have computed temperature-dependent effective
recombination rates for neutral oxygen in a temperature range
suitable for all types of SN nebulae (at epochs between 100 and 600 days). Since oxygen is the most
abundant element in stripped-envelope CC-SNe, oxygen lines are of
special interest among other recombination lines. 

We obtained core ejecta models for CC-SNe 1998bw and 2002ap. Similar
models had been derived previously.
Using these oxygen profiles, the O~{\sc i}~7774~\AA\
recombination line, which is the strongest observed
recombination line, is calculated and compared to observations. We show that up to late epochs
pure recombination is too weak to power O~{\sc i}~7774~\AA .
At earlier epochs the line is powered by scattering and
possibly by thermal electron excitation of recombining electrons. In
both scenarios the population of the 3s($^5$S) state by recombining
electrons is the key to the O~{\sc i}~7774~\AA\ emission.

We derived estimates for the strength of the O~{\sc i}~7774~\AA\ line resulting from oxygen
recombination, for the time-dependent optical depth of this
line and for excitation of recombining electrons by thermal collisions. These estimates give
insight
into the formation of the O~{\sc i}~7774~\AA\ line in CC-SNe. We have shown that while the recombination line strength depends on clumping
very weakly, clumping does have a strong influence on the absorption
and thermal electron excitation 
of this line.

Constraints on the clumping factor have been rare so far. A common
choice is a value of $\zeta \sim$ 10, consistent with the findings
of this paper. At the current level of
accuracy it is only possible to set upper and lower limits to the
clumping factor (100 $\gg \zeta > 1$).

Our results imply that the [O
  {\sc i}] 7774~\AA\ line should not be used as a tracer of the 
core ejecta before 250 days, unless one explicitly models the background radiation field and the
recombination and absorption processes involved. Since the clumping
factor can not be obtained with high accuracy and the background around 7774~\AA\ seems to consist of
a superposition of several weak lines this could become a difficult
task. It is not clear which elements or ions produce the observed
background flux in this
region. Iron group elements seem to be promising candidates, but
currently the accuracy of our calculations is not sufficient.  

\section*{Acknowledgments}
We thank Keith Butler for reading the manuscript and for making useful comments.

\bibliography{pap4}

\section*{Appendix A}

\begin{figure} 
\begin{center}
\includegraphics[width=8.5cm, clip]{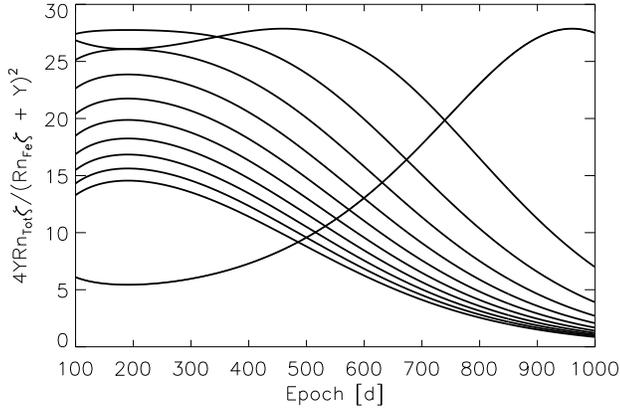}
\end{center}
\caption{The ratio
  $4Y'_\mathrm{O[I]}R_\mathrm{O[I]}n_\mathrm{Tot}\zeta/(R_\mathrm{O[I]}n_\mathrm{Fe}\zeta
    + Y'_\mathrm{O[I]})^2$ for different clumping factors ($\zeta =$ 1, 10, 20, 30, 40,
    50, 60, 70, 80, 90 and 100 from top to bottom) at 1000 days calculated for the
    same model as used in Figures \ref{pap4test4},\ref{pap4test1} and
    \ref{pap4test2}. For this model the ratio is much larger than one for any reasonable
    clumping factor (1 - 100) at epochs earlier than 600 days.} 
\label{pap4test3}
\end{figure}

\begin{figure} 
\begin{center}
\includegraphics[width=8.5cm, clip]{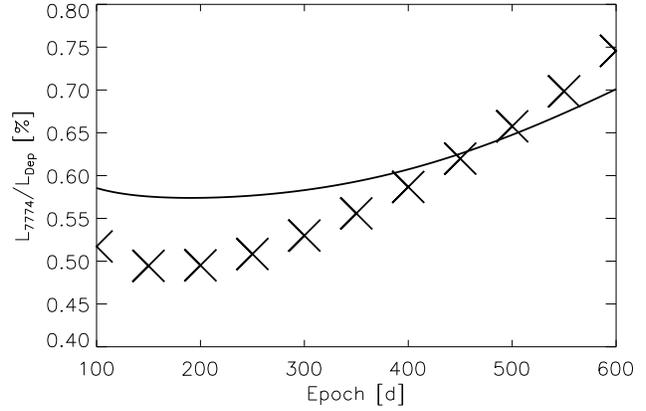}
\end{center}
\caption{The solid line is a computation of Equation \ref{Lx}
  performed for the O {\sc i} 7774 \AA\ line assuming a constant
  temperature of 4000~K at all epochs for the one-zone model described in
  the text. The luminosity of this line is compared
  to the total deposited luminosity. The results are given in $\%$. The recycling fraction was set
  to the ratio of ground state to total recombination rate at 4000~K. The crosses show data points
  obtained with the nebular code for the same model, where the
  temperature is computed at all epochs self-consistently and
ionisation is treated in detail. Considering the
large change of the temperature between 100 and 600 days (6800~K and
2800~K respectively) the
agreement seems reasonable (note that the temperature calculated by
the code drops below 4000~K around 450 days). The minimum around 200 days is present in
both curves.}
\label{pap4test1}
\end{figure}

\begin{figure} 
\begin{center}
\includegraphics[width=8.5cm, clip]{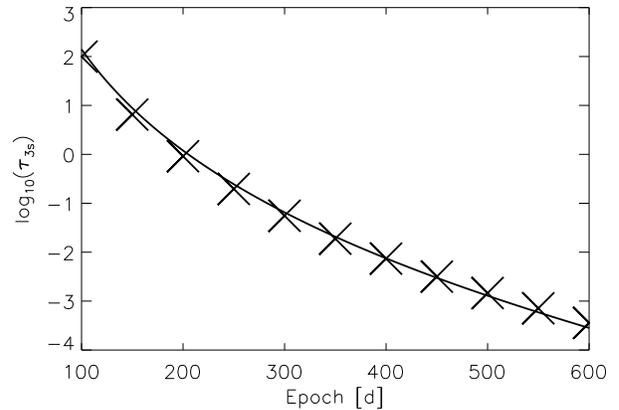}
\end{center}
\caption{The solid line shows a computation of Equation \ref{tau6}
  performed for the logarithmic line opacity of the O {\sc i} 7774 \AA\ line
  assuming a constant temperature of 4000~K at all epochs for a
  one-zone model described in the text. The crosses show data points
  obtained with the nebular code for the same model, where
the temperature is computed at all epochs self-consistently and
ionisation is treated in detail. The agreement is good, especially
around 450 days, when the simulated temperature drops below 4000~K,
which we assumed in our calculation. The line optical depth is greater
than one until day $\sim$ 200.}
\label{pap4test2}
\end{figure}

\begin{figure} 
\begin{center}
\includegraphics[width=8.5cm, clip]{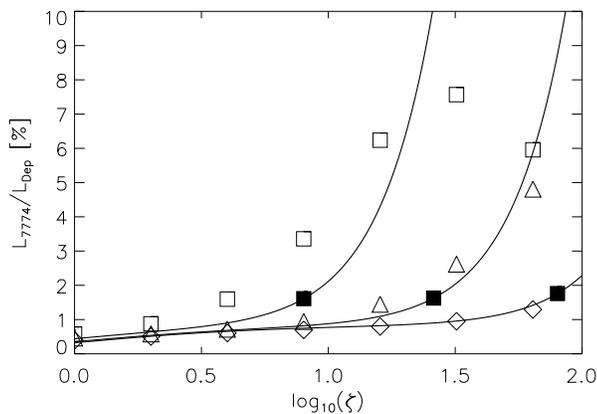}
\end{center}
\caption{The O {\sc i} 7774~\AA\ line luminosity from pure
  recombination and thermal excitation of recombining electrons relative to the total
deposited luminosity in \% for the test models described in the text
using clumping factors between 1 and 100. Solid lines show estimates obtained from Equations \ref{Lx} and
\ref{L7} for oxygen densities of 10$^7$, 10$^{7.5}$ and 10$^8$
cm$^{-3}$ from bottom to top. Results from our nebular code are shown
by diamonds (10$^7$ cm$^{-3}$), triangles (10$^{7.5}$ cm$^{-3}$) and
squares (10$^8$ cm$^{-3}$). The agreement is good. Small deviations are predominately caused by our rough approximation of the
temperature. Filled squares mark
$n_\mathrm{O}\zeta \sim 10^9$ cm$^{-3}$ where our estimates are expected
to become inaccurate. At $n_\mathrm{O}\zeta >
10^{9.3}$ cm$^{-3}$ our approximation breaks down completely, but
this regime is probably not important for the nebular phase of stripped CC-SNe. The
7774~\AA\ luminosity increases with the increasing clumping until the
density becomes high enough to depopulate the 3s($^5$S) state by
thermal electron collisions effectively. }
\label{pap4test6}
\end{figure}

In Section \ref{reclf} and \ref{oxex} we have derived several estimates for quantities
relevant for calculating O {\sc i} 7774~\AA\ nebular emission. To derive
these estimates, we used some approximations. These are
justified here by comparison to results from the nebular code, which
are not based on these approximations and will show deviations as soon as
our approximations fail. An important difference is introduced
by the temperature. While the electron temperature is calculated considering
hundreds of emission lines in the nebular code, this is not possible
for our estimates and we use a temperature of 4000~K for the calculations.

To perform these tests we set up a test one-zone model with a
total mass M$_\mathrm{Tot} \sim 0.5$ M$_\odot$,
an expansion velocity of 5000~km/s, containing 80$\%$ oxygen, 10$\%$ $^{56}$Ni,
$\sim$ 9$\%$ carbon, 
 and other elements like calcium, magnesium and sodium.

Usually one can assume
4$Y'_\mathrm{O[I]}R_\mathrm{O[I]}n_\mathrm{Tot}\zeta/(R_\mathrm{O[I]}n_\mathrm{Fe}\zeta
+ Y'_\mathrm{O[I]})^2  \gg $ 1, which depends
on the iron to oxygen ratio, the deposited luminosity, the
electron temperature and the clumping factor. In
Figure \ref{pap4test3} we show this ratio for our test model using
clumping factors of one to one hundred. At any epoch of interest (100
to 600 days) and for any clumping factor the assumption is valid. The
assumption may fail for clumping factors $\sim$ 1 at lower densities
and for higher $^{56}$Ni to oxygen
ratios. For clumping factors $\zeta \gg$ 1 the assumption will hold for
any plausible CC-SN scenario. 

In Figures \ref{pap4test1} and \ref{pap4test2} we compare Equations
\ref{Lx} and \ref{tau6} (temperature
set to T = 4000~K) with the results obtained
from our nebular code using our test model. The temperature is kept constant in the calculation,
while it is calculated self-consistently in the nebular code, influencing the recombination fraction and
recombination rate.  Also the ionisation of ions other than O {\sc i}
and Fe {\sc i} has been neglected in the calculation. Despite the
approximations, the agreement seems reasonable. 

In Figure \ref{pap4test6} we compare the O {\sc i} 7774~\AA\ flux from pure recombination and electron
scattering of recombining electrons (Equations \ref{Lx} and \ref{L7})
to results obtained from our nebular code using variations of our test
model. We set the
epoch to 150 days and use
oxygen densities of 10$^7$, 10$^{7.5}$ and 10$^8$ cm$^{-3}$, which
should cover the range expected for stripped CC-SNe. The oxygen fraction
is again set to 80\%. For our estimates
we use a temperature of 5000~K (the temperature calculated
by the nebular code varies between 4500~K and 6300~K for these models). The breakdown
point of our estimate ($n_\mathrm{O}\zeta > 10^9$ cm$^{-3}$) is indicated by
filled squares. Up to the breakdown point the agreement is good. For
larger clumping factors the agreement is acceptable for
$n_\mathrm{O}\zeta < 10^{9.3}$ cm$^{-3}$. For larger clumping factors our
estimate over predicts the 7774~\AA\ luminosity. In this regime the
population of the 3s($^5$S) level is no longer controlled by the
effective radiative rate but rather by collisional de-excitation by thermal
electrons. It seems unlikely that such high density regimes can play
a role in CC-SN nebulae.

\section*{Appendix B}

\begin{table*}
 \centering
 \begin{minipage}{140mm}
  \caption{Effective normalised recombination fraction of O {\sc i}
    for triplet and quinted
states of O {\sc i} n = 3. The rates are normalised on the total
radiative recombination rate of \citet{Badnell06}. States indicated by
$^*$ show the case of optical thin
ground state transitions. Values in brackets indicate a power of 10,
e.g. a(b)~=~a~$\cdot$~10$^b$.}
  \begin{tabular}{ccccccccccc}
  \hline
          State & 1000K &2000K & 3000K & 4000K & 5000K & 6000K & 7000K & 8000K & 9000K& 10000K  \\
 \hline
    3s($^3$S)$^*$ &4.6(-2)&4.2(-2)&4.0(-2)&3.8(-2)&3.7(-2)&3.6(-2)&3.4(-2)&3.3(-2)&3.2(-2)&3.1(-2)\\
    3p($^3$P)$^*$ &4.3(-2)&4.0(-2)&3.8(-2)&3.6(-2)&3.5(-2)&3.4(-2)&3.2(-2)&3.0(-2)&2.9(-2)&2.8(-2)\\
    3d($^3$D)$^*$ &1.6(-1)&1.4(-2)&1.3(-1)&1.2(-1)&1.1(-1)&1.1(-1)&1.0(-1)&9.4(-2)&8.9(-2)&8.5(-2)\\
\hline
    3s($^3$S)   &2.7(-1)&2.5(-1)&2.4(-1)&2.3(-1)&2.2(-1)&2.1(-1)&2.0(-1)&1.9(-1)&1.8(-1)&1.7(-1)\\
    3p($^3$P)   &2.6(-1)&2.4(-1)&2.3(-1)&2.1(-1)&2.0(-1)&2.0(-1)&1.9(-1)&1.8(-1)&1.7(-1)&1.6(-1)\\
    3d($^3$D)   &1.7(-1)&1.5(-1)&1.4(-1)&1.3(-1)&1.2(-1)&1.2(-1)&1.1(-1)&1.0(-1)&9.8(-2)&9.3(-2)\\
\hline
    3s($^5$S)   &4.7(-1)&4.5(-1)&4.3(-1)&4.1(-1)&3.9(-1)&3.8(-1)&3.6(-1)&3.5(-1)&3.3(-1)&3.2(-1) \\
    3p($^5$P)   &4.6(-1)&4.3(-1)&4.1(-1)&3.9(-1)&3.8(-1)&3.6(-1)&3.5(-1)&3.3(-1)&3.2(-1)&3.0(-1) \\
    3d($^5$D)   &2.7(-1)&2.4(-1)&2.2(-1)&2.1(-1)&1.9(-1)&1.8(-1)&1.7(-1)&1.6(-1)&1.5(-1)&1.5(-1) \\
\hline
 \hline
    4s($^3$S)$^*$ &6.9(-3)&7.0(-3)&7.0(-3)&6.8(-3)&6.6(-3)&6.4(-3)&6.2(-3)&6.0(-3)&5.7(-3)&5.5(-3)\\
    4p($^3$P)$^*$ &9.0(-3)&9.1(-3)&9.0(-3)&8.7(-3)&8.4(-3)&8.1(-3)&7.8(-3)&7.5(-3)&7.1(-3)&6.8(-3)\\
    4d($^3$D)$^*$ &4.2(-2)&4.0(-2)&3.8(-2)&3.7(-2)&3.5(-2)&3.3(-2)&3.1(-2)&2.9(-2)&2.8(-2)&2.6(-2)\\
    4d($^3$F)$^*$ &8.5(-2)&6.8(-2)&5.9(-2)&5.1(-2)&4.6(-2)&4.1(-2)&3.8(-2)&3.4(-2)&3.1(-2)&2.8(-2)\\
\hline
    4s($^3$S)   &3.8(-2)&3.8(-2)&3.7(-2)&3.6(-2)&3.4(-2)&3.3(-2)&3.2(-2)&3.0(-2)&2.9(-2)&2.7(-2)\\
    4p($^3$P)   &5.2(-2)&5.2(-2)&5.1(-2)&4.9(-2)&4.7(-2)&4.5(-2)&4.3(-2)&4.1(-2)&3.9(-2)&3.6(-2)\\
    4d($^3$D)   &4.5(-2)&4.4(-2)&4.2(-2)&4.0(-2)&3.8(-2)&3.6(-2)&3.4(-2)&3.2(-2)&3.0(-2)&2.9(-2)\\
    4d($^3$F)   &8.6(-2)&6.9(-2)&5.9(-2)&5.2(-2)&4.7(-2)&4.2(-2)&3.8(-2)&3.5(-2)&3.2(-2)&2.9(-2)\\
\hline
    4s($^5$S)   &6.8(-2)&6.9(-2)&6.7(-2)&6.5(-2)&6.3(-2)&6.1(-2)&5.8(-2)&5.5(-2)&5.3(-2)&5.0(-2) \\
    4p($^5$P)   &8.1(-2)&8.1(-2)&8.0(-2)&7.7(-2)&7.4(-2)&7.1(-2)&6.8(-2)&6.5(-2)&6.1(-2)&5.8(-2) \\
    4d($^5$D)   &7.1(-2)&6.8(-2)&6.5(-2)&6.2(-2)&5.9(-2)&5.6(-2)&5.3(-2)&5.0(-2)&4.7(-2)&4.4(-2) \\
    4d($^5$F)   &1.4(-1)&1.1(-1)&9.8(-2)&8.6(-2)&7.7(-2)&7.0(-2)&6.3(-2)&5.8(-2)&5.3(-2)&4.8(-2) \\
\hline
\end{tabular}
\end{minipage}
\end{table*}

Normalised effective recombination fractions for triplet and quinted
states of O {\sc i} n = 3. For the triplet states we show calculations for
the cases of optical thin $^*$ and thick ground states. These numbers have
to be multiplied by the total radiative recombination rate to obtain
the effective
recombination rates. Since the total radiative recombination rate also contains
a direct ground state component and since lower levels contain
cascade contributions from the higher levels these fractions do not
add up to one.

At high densities, collisional transitions become important, which are
not included in our cascade calculation. At densities $n_\mathrm{e}
< 10^{10}$ cm$^{-3}$ and temperatures of $T \sim 5000$K the error on
the recombination rates owing to this effect should be less than 10$\%$. 

At high temperatures di-electronic recombination becomes important,
while at low temperature and high densities collisional recombination
can become important. Under nebular conditions both effects should be
of the order of 10\% or less.

To obtain the effective recombination coefficient for a certain line
one has to weigh the upper level by the corresponding radiative
rates\footnote{http://www.nist.gov/index.html}. As an example, we calculate the normalised effective recombination
coefficient $f_{6157}$ of the quinted 4d($^5$D) $\rightarrow$ 3p($^5$P) 6157
\AA\ line at 1000K. The radiative rate from 4d($^5$D) to 3p($^5$P) is
given by $A_{6157}$~=~7.62(+6)~s$^{-1}$ and the radiative rate 4d($^5$D) to 4p($^5$P)
is given by $A_{26511} = 6.44(+6)$~s$^{-1}$ and therefore 
\begin{equation}
f_{6157} = f_\mathrm{4d(^5D)}\frac{A_{6157}}{A_{6157} + A_{26511}} = 3.8(-2)
\end{equation}
which agrees well with the value given by \citet{Julienne74} at 1160K.

\end{document}